\documentclass[atmp]{ipart_v1}
\usepackage{amssymb}
\usepackage{amsfonts}
\usepackage{graphicx}
\usepackage{epstopdf}
\usepackage{dcolumn}
\usepackage{amsmath}
\usepackage{latexsym,bm}
\usepackage{amsthm}
\usepackage{slashed}
\usepackage{float}

\def \be {\begin{equation}}
\def \ee {\end{equation}}
\def \bsp {\begin{split}}
\def \esp {\end{split}}
\def \bea {\begin{eqnarray}}
\def \eea {\end{eqnarray}}

\def\mc{\mathcal}
\def\cal{\mathcal}

\def\ge{{\mathfrak{e}}}
\def\gso{{\mathfrak{so}}}
\def\gsu{{\mathfrak{su}}}
\def\gsp{{\mathfrak{sp}}}
\def\gf{{\mathfrak{f}}}
\def\gg{{\mathfrak{g}}}

\def\P{\mathbb{P}}
\def\C{\mathbb{C}}
\def\Z{\mathbb{Z}}
\def\F{\mathbb{F}}
\def\Q{\mathbb{Q}}

\newtheorem{lemma}{Lemma}
\newtheorem{prop}{Proposition}
\newcommand{\fracs}[2]{{\textstyle{#1\over #2}}}

\begin{document}

\title{Non-toric bases for elliptic Calabi-Yau
threefolds and 6D F-theory vacua}

\author[Washington Taylor, Yi-Nan Wang]{Washington Taylor, Yi-Nan Wang}

\rightline{MIT-CTP-4629}

\begin{abstract}

We develop a combinatorial approach to the construction of
  general smooth compact base surfaces that support elliptic Calabi-Yau
  threefolds.  This extends previous analyses that have relied on
  toric or semi-toric structure.  The resulting algorithm is used to
  construct all classes of such base surfaces
  $S$ with $h^{1, 1} (S) < 8$ and all base surfaces over which there
  is an elliptically fibered Calabi-Yau threefold $X$ with Hodge
  number $h^{2, 1} (X) \geq 150$.  These two sets can be used to
  describe all 6D F-theory models that have fewer than seven tensor
  multiplets or more than 150 neutral scalar fields respectively in
  their maximally Higgsed phase.  
  Technical challenges to constructing the complete list of base
  surfaces for all Hodge numbers are discussed. 
\end{abstract}

\maketitle

\section{Introduction}

The problem of classifying Calabi-Yau threefolds is important both for
mathematics and for physics.  In mathematics, progress in recent years
on the minimal model program for surfaces and for higher-dimensional
algebraic varieties ({\it i.e.}  Mori theory \cite{Mori}) has provided
tools for understanding and classifying threefolds and
higher-dimensional varieties with non-vanishing canonical class.
These methods are not directly applicable, however, to classification
of Calabi-Yau manifolds, which have a canonical class that vanishes
(up to torsion).  In physics, Calabi-Yau manifolds play an important
role as compactification spaces for string theory \cite{chsw}.
Classification of Calabi-Yau manifolds is useful in understanding the
space of vacuum solutions of string theory.  Despite many years of
study of these spaces, it is still not known if the number of distinct
topological types of Calabi-Yau threefolds is finite or infinite.
Physicists have systematically analyzed certain types of Calabi-Yau
manifolds using specific constructions such as complete intersections
in projective spaces (CICY's \cite{CICY}) or hypersurfaces in toric
varieties \cite{Batyrev}.  A particularly useful compilation of data
for the full set of toric hypersurface Calabi-Yau's was made by
Kreuzer and Skarke \cite{Kreuzer-Skarke} (see also \cite{CY-data-2}
for more refined data for the threefolds with small $h^{1, 1}$, and
\cite{CY-explorer} for a nice graphical interface to the set of
current Calabi-Yau data.).  See
\cite{Davies, He} for recent reviews of various approaches to
constructing and classifying Calabi-Yau threefolds.

The space of \emph{elliptically fibered} Calabi-Yau threefolds represents a
subset of the set of Calabi-Yau manifolds that is known to admit a
finite number of topological types \cite{Gross}.  In recent years,
motivated by the physics of F-theory \cite{F-theory, Morrison-Vafa-I,
  Morrison-Vafa-II}, a systematic approach has been developed to
classifying elliptically fibered Calabi-Yau threefolds through the
geometry of the base of the elliptic fibration, which is a complex
surface.  The set of possible configurations of mutually intersecting
curves of self-intersection $-2$ or below that can arise on a surface
supporting an elliptic fibration where the total space is Calabi-Yau
were classified in \cite{Classbasis}.  These ``non-Higgsable
clusters'' provide building blocks out of which all base surfaces that
support elliptic Calabi-Yau threefolds can be constructed.  
The
complete set of allowed compact toric bases was described and enumerated in
\cite{Toric}; this analysis was extended to bases with a single
$\C^*$-structure in \cite{semi-toric}.  For each allowed base, there
are many possible different elliptic Calabi-Yau manifolds over that
base; each distinct topology can be realized by resolving singular
geometries constructed by ``tuning'' coefficients in a generic
Weierstrass model over the given base (see,
{\it e.g.}, \cite{JohnsonTaylor}) to realize
different Kodaira singularity types over divisors in the base.
Non-Higgsable clusters have also been used as basic building blocks in
the construction of general 6D superconformal field theories
\cite{SCFT-1, SCFT-2, SCFT-3}.

In this paper we develop methodology to  classify general smooth compact
bases that support elliptic Calabi-Yau threefolds, without restricting
to base surfaces with toric or any other specific structure.  The approach we
take is combinatorial in nature.  From the minimal model program for
surfaces \cite{bhpv} and the work of Grassi \cite{Grassi} it is known
that all allowed bases can be formed from blow-ups of the Hirzebruch
surfaces $\F_m, m \in\{0, 2,3,  \ldots, 7, 8, 12\}$, projective space $\P^2$,
or the Enriques surface.  Our basic approach is to characterize each
surface in terms of the combinatorics of the cone of effective
curves.  This data can be characterized by the set of vectors that
generate the effective cone in the
lattice $\Z^{1, T}$, where
$T = h^{1, 1} (S) -1$
corresponds to the number of tensor multiplets in an F-theory
compactification on the base surface $S$.  In most cases,
the cone has a finite number of generators and this is a finite
combinatorial problem.

There are a number of technical complications that can
arise in this
analysis.  In some cases, the number of generators of the effective
cone can become infinite.  In other situations, the combinatorial
structure of the cone of effective curves is insufficient to uniquely
specify the geometry of the base surface.  In this paper we describe
these complications and develop an algorithm  for explicitly
constructing bases that avoids these issues in some
regions of interest.  This gives a systematic approach to constructing
large classes of bases for elliptic Calabi-Yau threefolds.  In
particular, we construct  bases with small $T$, and  bases that
give generic elliptic fibrations corresponding to Calabi-Yau
threefolds with Hodge number $h^{2, 1}(X) \geq 150$.  In both cases,
we believe that we have constructed all possible bases in the
appropriate regime, though we have not proven this in a completely
rigorous mathematical fashion.  In principle, a
  systematic analysis of tuning Weierstrass moduli over these bases
  could lead to a systematic construction of all elliptic Calabi-Yau
  threefolds with large $h^{2, 1}(X)$ or small $h^{1, 1} (S)$.

In \S\ref{sec:F-theory-bases}, we review the basics of F-theory and
the connections between the geometry of a complex base surface $S$, the
generic elliptic Calabi-Yau threefold $X$ over $S$, and the physics of
the corresponding F-theory compactification.  We also review some
basic complex algebraic geometry of surfaces.  In \S\ref{sec:example},
we give some very simple examples of base surfaces and the structure
of the cone of effective curves on these bases.  In
\S\ref{sec:curves}, we describe in some detail how the cone of
effective curves can be determined when a surface is formed as the
blow-up of another surface.  We describe a more general class of
examples, the generalized del Pezzo surfaces, in \S\ref{sec:gdp}.  In
\S\ref{sec:obstructions} and \S\ref{sec:finite}, we discuss
obstructions to the systematic combinatorial construction of all base
surfaces.  In \S\ref{sec:algorithm} we describe the systematic
algorithm for classification of bases, and give the results of
applying the algorithm for small $T$ in \S\ref{sec:small} and for
large $h^{2, 1} (X)$ in \S\ref{sec:large}.  Some concluding remarks
are given in \S\ref{sec:conclusions}

Note that while the analysis  in this paper is carried out
in a concrete mathematical framework, we have not attempted to be
completely rigorous from the mathematical point of view.  The goal of
this work is to explore the space of general bases that support
elliptic CY threefolds using a combinatorial and algorithmic approach
that gives insight into the nature of the scope and structure of
generic bases of this type.  In various places in the paper we point
out some specific ways in which our analysis neglects certain
complexities which would need to be addressed more systematically for
a complete classification of all bases supporting elliptic CY
threefolds with arbitrary Hodge numbers.

\section{Complex surfaces as F-theory bases}
\label{sec:F-theory-bases}

We summarize here some basic facts from the physics of F-theory and
the algebraic geometry of complex surfaces.  For a more detailed
introduction to F-theory, see the reviews \cite{Morrison-TASI,
  Denef-F-theory, WT-TASI}.

\subsection{F-theory on elliptic Calabi-Yau threefolds}

When F-theory is compactified on a compact Calabi-Yau threefold $X$
that is elliptically fibered, one gets a 6D theory of
gravity with $\mathcal{N}=1$ supersymmetry.  This theory may also
contain tensor, vector, and scalar (hypermultiplet) fields.

The elliptic fibration $\pi: X \rightarrow S$ must
have a global section;
a Weierstrass model of any such fibration exists \cite{Nakayama}
and takes the form
\be
y^2=x^3+fx+g.\label{Weierstrass}
\ee
If we denote the canonical class of the base by $K_S$, then the
functions $f$, $g$ and $\Delta$ are sections of line bundles ${\cal O}(-4K_S)$,
${\cal O}(-6K_S)$ and ${\cal O} (-12K_S)$ respectively.

The fiber is singular along loci in the base manifold $S$ where the
discriminant
\be
\Delta=4f^3+27g^2
\ee
vanishes.
The Kodaira-Tate classification characterizes codimension-one
singularities where 
the discriminant $\Delta$ vanishes 
\cite{Kodaira, Tate}. When the
functions $f$, $g$ and $\Delta$ vanish to specific orders
on a divisor $D$,
the singularity is associated with the Dynkin diagram of a
certain non-Abelian gauge algebra that describes a gauge field in the
corresponding physical 6D supergravity theory
\cite{Morrison-Vafa-I, Morrison-Vafa-II, Bershadsky-all, Morrison-sn,
  Grassi-Morrison-2}.  The correspondence between vanishing orders of
$f, g$ and nonabelian symmetry algebras is listed in Table
\ref{t:Kodaira}.
Codimension two singularities give rise to matter fields in the 6D
theory, though this correspondence is not yet completely classified.
Note that if $(f,g)$ vanish to order $(4,6)$ or higher at
a codimension-one locus on the base,
the singular Weierstrass model (\ref{Weierstrass}) cannot be resolved into a
smooth Calabi-Yau threefold. Then the F-theory compactification does not have a supersymmetric
M-theory dual, hence it will not give any supergravity vacua.
These points are also at infinite distance in the Weierstrass moduli space.
If $(f,g)$ vanishes to order $(4,6)$ or higher at a codimension-two locus (points) on
 the base, then such points need to be blown up on the base in the
 resolution process to give a smooth Calabi-Yau threefold. 

\begin{table}
\begin{center}
\begin{tabular}{|c |c |c |c |c |c |}
\hline
Type &
ord ($f$) &
ord ($g$) &
ord ($\Delta$) &
sing. &  symmetry algebra\\ \hline \hline
$I_0$&$\geq $ 0 & $\geq $ 0 & 0 & none & none \\
$I_n$ &0 & 0 & $n \geq 2$ & $A_{n-1}$ & $\gsu(n)$  or $\gsp(\lfloor
n/2\rfloor)$\\
$II$ & $\geq 1$ & 1 & 2 & none & none \\
$III$ &1 & $\geq 2$ &3 & $A_1$ & $\gsu(2)$ \\
$IV$ & $\geq 2$ & 2 & 4 & $A_2$ & $\gsu(3)$  or $\gsu(2)$\\
$I_0^*$&
$\geq 2$ & $\geq 3$ & $6$ &$D_{4}$ & $\gso(8)$ or $\gso(7)$ or $\gg_2$ \\
$I_n^*$&
2 & 3 & $n \geq 7$ & $D_{n -2}$ & $\gso(2n-4)$  or $\gso(2n -5)$ \\
$IV^*$& $\geq 3$ & 4 & 8 & $\ge_6$ & $\ge_6$  or $\gf_4$\\
$III^*$&3 & $\geq 5$ & 9 & $\ge_7$ & $\ge_7$ \\
$II^*$& $\geq 4$ & 5 & 10 & $\ge_8$ & $\ge_8$ \\
\hline
non-min &$\geq 4$ & $\geq6$ & $\geq12$ & \multicolumn{2}{c|}{does not
appear for SUSY vacua} \\
\hline
\end{tabular}
\end{center}
\caption[x]{\footnotesize  Table of
codimension one
singularity types for elliptic
fibrations and associated nonabelian symmetry algebras.
In cases where the algebra is not determined uniquely by the degrees
of vanishing of $f, g$,
the precise gauge algebra is fixed by monodromy conditions that can be
identified from the form of the Weierstrass model.
}
\label{t:Kodaira}
\end{table}

Rational curves of negative self-intersection on the base surface $S$
can enforce minimal degrees of vanishing of $f$, $g$, $\Delta$ over
those curves, for a generic elliptic fibration.  When this happens,
the physical theory has some minimal content of gauge fields and
matter fields, regardless of the specific fibration one chooses in the
Weierstrass moduli space.  The configurations of curves of negative
self-intersection that give rise to gauge and matter fields in this
way were classified in \cite{Classbasis}, and dubbed ``non-Higgsable
clusters'' (NHCs), since the 6D gauge groups that arise in this way
cannot be broken in a fashion that preserves supersymmetry by giving
vacuum expectation values to ({\it i.e.}, ``Higgsing'') the charged
matter fields. We have listed the configurations of possible NHCs in
Table \ref{t:clusters}.  While the gauge groups produced by
non-Higgsable clusters cannot be broken in any supersymmetric vacuum
without changing the base $S$, the coefficients of the Weierstrass
model, associated with neutral scalar fields in the 6D supergravity
theory, may be tuned to enhance the gauge contents of the theory.  It
was shown, for example, that in this way all elliptically fibered
Calabi-Yau threefolds with $h^{2, 1}(X)\geq 350$ (and indeed all known
Calabi-Yau threefolds exceeding this Hodge number bound) can be
constructed through tuning Weierstrass models over simple complex base surfaces
\cite{JohnsonTaylor}.

\begin{table}
\begin{center}
\begin{tabular}{| c |
c | c |c |c |
}
\hline
Cluster & gauge algebra & $R_{\rm nonabelian}$ & $V$ &  $H_{\rm charged}$
\\
\hline
(-12) &$\ge_8$ & 8 & 248 &0 \\
(-8) &$\ge_7$&  7 &  133 &0 \\
(-7) &$\ge_7$& 7 & 133 &28 \\
(-6) &$\ge_6$&   6 & 78 &0 \\
(-5) &$\gf_4$&   4 & 52 &0 \\
(-4) &$\gso(8) $&  4 & 28 &0 \\
(-3, -2, -2)  &  $\gg_2 \oplus \gsu(2)$&  3 & 17 &8\\
(-3, -2) &  $\gg_2 \oplus \gsu(2)$ &3 & 17 &8 \\
(-3)& $\gsu(3)$ &  2 & 8 &0 \\
(-2, -3, -2) &$\gsu(2) \oplus \gso(7) \oplus
\gsu(2)$&5 & 27 &16 \\
(-2, -2, \ldots, -2) & no gauge group & 0 & 0 &0 \\
\hline
\end{tabular}
\end{center}
\caption[x]{\footnotesize
List of ``non-Higgsable clusters'' of
  irreducible effective divisors with self-intersection $-2$ or below,
  and corresponding contributions to the gauge algebra and matter
  content of the 6D theory associated with F-theory compactifications
  on a generic elliptic fibration (with section) over a base
  containing each cluster.
The quantities $R_{\rm nonabelian}$ and $V$ denote the rank and dimension of the
nonabelian gauge algebra, and $H_{\rm charged}$ denotes the number of
charged hypermultiplet matter fields, which are in all cases other
than the $-7$ curve cluster associated with intersections
between the curves supporting the gauge group factors.
}
\label{t:clusters}
\end{table}

The Hodge numbers of the generic ({\it i.e.}, untuned) elliptic Calabi-Yau
threefold $X$ associated to a given base
$S$ can be easily computed
from the non-Higgsable cluster content of the base
\cite{WT-Hodge}
using the relations
\bea
&&h^{1,1}(X)=R+T+2\label{eq:h11}\\
&&h^{2,1}(X)=272-29T+V-H_{\mathrm{charged}}. \label{eq:h21}
\eea
Here, as above, $T=h^{1,1}(S)-1$ is the number of tensor multiplets.
The quantity $R$ is the total rank of (nonabelian + abelian) gauge
groups.  $V$ and $H_{\mathrm{charged}}$ are the numbers of
6D vector supermultiplets
and charged matter hypermultiplets
respectively \cite{Morrison-Vafa-I, Morrison-Vafa-II}. $V$,
$H_{\mathrm{charged}}$, and the contribution of nonabelian gauge
groups to $R$ can be directly computed by summing up the
contributions of all NHCs, using Table \ref{t:clusters}, when the intersection structure of curves on the base
is known.
In principle, there can also  be an abelian contribution to $R$, even
when $X$ is the generic elliptic CY threefold over the base $S$.
While this does not occur for toric base surfaces $S$, it was found in
\cite{semi-toric} that there are a small number of base surfaces with
a single $\C^*$ structure (``semi-toric'' bases) that give rise to
theories with a non-Higgsable abelian gauge group structure.
The corresponding Calabi-Yau threefolds are elliptic fibrations with
multiple independent sections, corresponding to a higher rank
Mordell-Weil group.
It is shown in \cite{Morrison-Park-Taylor} that all these threefolds
are related and can be described through a generalization of the
Schoen construction \cite{Schoen}.
In general, computing the rank of the Mordell-Weil group for a given
compactification space is a difficult mathematical problem, which we
do not attempt to address here. 
For the specific classes of non-toric bases that we enumerate here,
there are no abelian contributions to $R$ for generic elliptic
fibrations.  For a comprehensive analysis of bases giving elliptic
Calabi-Yau threefolds with large $h^{1, 1}$
and small $h^{2, 1}$, however, this
possibility would need to be considered further.

As mentioned above, the base $S$ of any elliptically fibered Calabi-Yau
threefold $X$ can be constructed by blowing up a finite number of
points on $\mathbb{P}^2$, a Hirzebruch surface $\mathbb{F}_n(0\leq
n\leq 12)$ or the Enriques surface. The F-theory models in the case of
the Enriques surface are relatively trivial, and do not allow any
gauge field or matter content, since $-K_S$ vanishes (up to torsion).
In this paper we only consider $\mathbb{P}^2$ and $\mathbb{F}_n(0\leq
n\leq 12)$ \cite{Classbasis}. In Kodaira's classification of complex
surfaces (not to be confused with his classification of codimension
one singularities in elliptic fibers mentioned above), these are all
rational surfaces (birational to $\mathbb{P}^2$), with Kodaira
dimension $\kappa(S)=-\infty$ \cite{bhpv}.  The condition that
$-4K_S,-6K_S,-12K_S$ have global sections means that the
anticanonical class $-K_S$ is 
in the cone of effective divisors.  Stated more technically, the
complex surfaces we study in
this paper are anticanonical rational
surfaces \cite{Looijenga, Harbourne1997}, or those where the anticanonical
class is an effective $\mathbb{Q}$-divisor.
We focus in this paper only on smooth base surfaces $S$.

\subsection{Algebraic geometry of rational complex surfaces}

The general theory of complex algebraic surfaces can be found in
\cite{Griffith-Harris, bhpv}.  We review here a few basic definitions
and ideas that will be helpful in systematically generating base
surfaces.

A (Weil) {\em divisor} $D$ is a formal linear combination with integer
coefficients of irreducible algebraic curves on the algebraic surface
$S$. If the coefficients in the linear combination are all
non-negative, then the divisor is {\em effective}, $D \geq 0$.  A {\em
  $\Q$-divisor} is defined similarly to a Weil divisor, but with
coefficients in $\Q$ (Note that
Weil divisors and Cartier divisors coincide on
smooth varieties, which is the only situation that we consider
here). The {\em divisor class} $[D]$ is the homology class of $D$.
The set of divisor classes modulo linear equivalence on $S$ forms an additive group, which is
homomorphic to the {\em Picard group} Pic$(S)$ of line bundles on the
surface (which form a group under the tensor product operation).  For
effective divisors, the divisors in the same divisor class as a given
divisor $D$ are said to be {\em linearly equivalent}, and form a {\em
  linear system}, denoted $|D|$.

There is a bilinear operation $\cdot:$Pic$(S)\times
$Pic$(S)\rightarrow\mathbb{Z}$, which corresponds to the
{\em intersection number} of
two divisor classes. 
For two distinct divisor classes $[D_1],[D_2]$, we can pick out
one specific representative $D_1\in|D_1|,D_2\in|D_2|$
from each class; the intersection
number $[D_1]\cdot[D_2]$
then just coincides with the geometric number of intersection points
(with multiplicity)
between the two curves $D_1$ and $D_2$. The self-intersection of a
divisor
class $[D]$ is $[D]\cdot[D]$.
Note that the self-intersection of a curve can be negative, but only
when the curve is rigid and cannot be deformed; otherwise a small
deformation would give two distinct curves that would intersect at a
finite positive number of points.  (because of the complex structure,
all intersections between complex curves in a surface have the same
orientation, and all contribute positively to intersection numbers.)

For $\mathbb{P}^2$, the inequivalent divisor classes are those of the
form $n[H]$, where $n \in\Z$ and $[H]$ is the divisor class of the
hyperplane in $\mathbb{P}^2$; the hyperplane $[H]$ is just a line
$\mathbb{P}^1 \subset\P^2$ given by a linear equation, for example as
described by the algebraic equation $z = 0$ in homogeneous coordinates
$[x:y:z]$ on $\P^2$.  The linear system $|nH|$ corresponds to the
space of all degree-$n$ curves on $\mathbb{P}^2$.

The mathematical procedure of {\em blowing up} a point $p$ on a smooth
surface $S$ consists of replacing the point by a copy of $\P^1$, called
the {\em exceptional curve}, where the points on the $\P^1$ correspond
to all (complex) directions in which lines on the original surface can
pass through the original point.  After a blow-up process, a new
effective divisor class arises on the resulting blown up surface
$S'$, which is the class of the exceptional
curve $[E]$. 
This divisor class has self-intersection $-1$; the linear
system of the exceptional curve only contains one curve, hence it is
generally just called {\em the} exceptional curve.  
There is a birational map $\pi:S' \rightarrow S$, which is one-to-one
and onto everywhere except on the exceptional curve, where $\pi:E
\rightarrow p$ and $p$ is the point being blown up.
If there is some
representative curve $D$ of a certain effective divisor class $[D]$
that passes through the point $p$, with multiplicity $m$, then after
the blow-up process, a new divisor class
\be
[D']=[D]-m[E] 
\ee
appears as an effective divisor class on $S'$.
This is called the \emph{proper transform} of $D$.
Here we have
identified $[D]$ on $S'$ with $\pi^{-1} ([D])$.

For a given surface $S$ generated by blowing up $\mathbb{P}^2$ 
consecutively $r$
times, a convenient basis of Pic$(S)$ consists of the divisor
class of the  hyperplane $[H]$ on the original $\mathbb{P}^2$ and the
exceptional divisors $[E_1],[E_2],\dots,[E_r]$. The intersection
matrix on this basis is given by
\be
[H]\cdot[H]=1\ ,\ [H]\cdot[E_i]=0\ ,\ [E_i]\cdot [E_j]=-\delta_{ij}.
\label{eq:Hodge-basis}
\ee
Hence the Picard rank is rk(Pic$(S))=r+1$.

More generally, the Hodge index theorem states that for any surface
with rk(Pic$(S)$) $ = r + 1$, the signature of the intersection matrix is
$(1, r)$.  When $r > 1$, there is always a basis in which the
intersection product takes the form (\ref{eq:Hodge-basis}); for $r =
1$, there are surfaces where the intersection form has the structure
of the matrix
\begin{equation}
U =
\left(
\begin{matrix}
0 & 1\\
1 & 0\\
\end{matrix}
\right)
\end{equation}

The canonical divisor class of $\P^2$
is $[K_S] = -3[H]$, and after $r$ blow-ups is always in the form of
\be
[K_S]=-3[H]+\sum_{i=1}^r [E_i].
\ee
More generally, the canonical divisor class can always be taken to
have this form in a basis with intersection form (\ref{eq:Hodge-basis}).
In general,
if an irreducible divisor class on a surface formed from $r$
blow-ups on $\P^2$ can be decomposed into a sum of the
hyperplane and other divisor classes as $[D]=n[H]+\dots$, then it is the
divisor class of some degree $n$ curve on the original $\P^2$.

There is a useful genus formula that follows from the adjunction formula:
\be
[C]\cdot([K_S]+[C])=2g-2\label{RiemannRoch}
\ee
where $g$ is the genus of the curve $C$. If $g=0$, then $C$ is a rational
curve. If $g=1$, then $C$ is an elliptic curve. Note that the
anticanonical divisor class $[-K_S]$ is always elliptic.

The dimension of the linear system $|D|$ is related to the dimension
of the cohomology class of sheaves $\mc{O}([D])$:
\be
\mathrm{dim}(|D|)=h^0(S,\mc{O}([D]))-1
\ee
For rational curves $D$, we have the following result from the Riemann-Roch theorem \cite{Kollar, CaporasoHarris,
  CaporasoHarris2}
\be
\mathrm{dim}(|D|)=([D]\cdot[D]-[K_S]\cdot [D])/2=[D]\cdot[D]+1.
\ee

If $\mathrm{dim}(|D|)>0$, then there exists a curve $D\in|D|$ that
passes through $\mathrm{dim}(|D|)$ general points. This implies
furthermore that if $[D]\cdot[C]>0$ for another divisor class $[C]$,
  then for every generic point $p$ on $C\in|C|$, there exists a curve
  $D\in|D|$ that passes through it.  This is the case whenever the
  class of the divisor $D$ has non-negative self-intersection. When the class
$[D]$ has negative self-intersection, then
  $\mathrm{dim}(|D|)\leq0$, which means that the curve in $|D|$ cannot
  be deformed.  In this case, if $[D]$ is irreducible, then it is
  really a single ``fixed'' curve.

For general (arbitrary genus) curves,
the value of $h^0(S,\mc{O}([D]))$ satisfies 
$h^0(S,{\cal O}(D))\geq([D]^2-[K_S]\cdot [D])/2+1$
 \cite{Griffith-Harris}.  

In general we have the following lemma:
\begin{lemma}
A negative divisor class on $S$ is always rigid. For a non-negative
effective divisor class $D$ on $S$, there exists a representative
$D\in|D|$ that passes through any $([D]^2-[K_S]\cdot [D])/2$
points in $S$.\label{l:passthrough}
\end{lemma}

More generally, this lemma also guarantees that curves of sufficiently
positive self-intersection can pass through points with arbitrary
multiplicity.  When $([D]^2-[K_S]\cdot [D])/2\geq 3$, for example,
there is a representative curve $D$ that passes through any point $p$ in
$S$ with multiplicity 2 (which requires two additional conditions,
corresponding to the vanishing of the two first derivatives at $p$).
In general, for a curve to exist that passes through any point with multiplicity
$m$, we must have  $([D]^2-[K_S]\cdot [D])/2\geq m(m+1)/2$. 

Note that the above statements are completely consistent with the
intuition of planar geometry. For example, consider the curves of
self-intersection $(-1)$ on a surface formed by blowing up $\P^2$ $r$ times.
The divisor classes of degree
1 rational curves with self-intersection $(-1)$ can be written as
$[H]-[E_i]-[E_j]$, corresponding to the set of lines that pass through
the pairs of
points $p_i$ and $p_j$ on the original $\mathbb{P}^2$. There is a
unique line that passes through two fixed points in the plane, hence
the resulting $(-1)$-curve is fixed. 
Similarly,
the divisor class of degree 2 rational curves with self-intersection
$(-1)$ can be written as $2[H]-[E_i]-[E_j]-[E_k]-[E_l]-[E_m]$,
corresponding to the set of conics that
pass through five fixed points. 
We know this divisor is also  rigid, since there is a unique conic
passing through five fixed points. 
If we consider, however, the divisor class of degree 1 rational curves
with self-intersection 0, $[H]-[E_i]$, these correspond to the set of lines that
pass through a particular point
$p_i$, and are free to move.
In fact, such lines can pass through
through every point on $\mathbb{P}^2$.

\begin{figure}
\centering
\includegraphics[height=3cm]{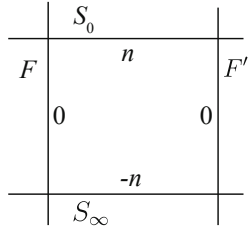}
\caption{Loop of irreducible effective curves on the Hirzebruch
  surface $\mathbb{F}_n$, corresponding to irreducible toric divisors
  associated with rays in the toric fan. $[F] = [F']$ correspond to
  the same divisor class, and $[S_0] =[S_\infty] + n F$. The
  self-intersections of each curve are labeled beside the divisor
  class}\label{f:Hirzebruch}
\end{figure}

We have focused so far on blow-ups of $\P^2$, but the above properties
of divisors can also be applied to the Hirzebruch surfaces $\F_n$
(Note that $\F_1$ is the surface that results from blowing up a single
point on $\mathbb{P}^2$). In the language of toric geometry, the
irreducible toric divisors form a cyclic diagram (see Figure
\ref{f:Hirzebruch}). For the cases with odd $n=2k-1$, the relevant
divisor classes can be written in the basis (\ref{eq:Hodge-basis})
using the following linear combinations of the two generators of
Pic$(\F_{2k-1})$:
\be
[S_0]=k
[H]-(k-1)[E_1]\ ,\ [S_\infty]:(-k+1)[H]+k[E_1]\ ,\ [F],[F']=[H]-[E_1].\label{Fn} 
\ee
The canonical class is, for all $k$,
\begin{equation}
K= -[S_0]-[S_\infty] -2[F] = -3[H] +[E_1] \,.
\end{equation}

The blowing-up of $\F_{2k-1}$ at $r-1$ points has the same Picard
groups and intersection matrices for each $k$, and matches with that
of $\P^2$ blown up at $r$ points.  Also, the canonical class is always
again in the form $K =-3[H]+\sum_{i=1}^r [E_i]$.  Generally speaking,
therefore, although the geometric intuition comes most easily from
$\mathbb{P}^2$,  the structure of divisors and intersection structure
is essentially the same on other rational surfaces arising as blowups of Hirzebruch
surfaces.

\begin{figure}[H]
\centering
\includegraphics[height=3.5cm]{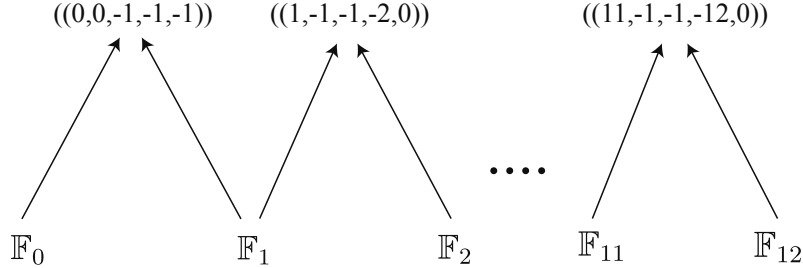}
\caption{Blowing up Hirzebruch surfaces.
For each surface
  $\mathbb{F}_{n\geq 0}$, there are only two different ways to blow up it:
blow up on the curve of negative self-intersection or blow up at a
generic point. For $\mathbb{F}_{12}$, the
(-12)-curve cannot be blown up, as the base of an elliptically fibered
Calabi-Yau threefold is not allowed to have curves with
self-intersection lower than $-12$ \cite{Classbasis}. 
The results of blowing up are represented as the self-intersections of
the (cyclic) sequence of toric divisors in the blown up surface.}\label{f:Chain}
\end{figure}

For even $n$, we can relate any blow-up of the Hirzebruch surface
$\F_n$ to the cases with odd $n$ via the blow-up chains shown in
Figure \ref{f:Chain}. Thus, every surface that results from blowing up
$\mathbb{F}_{2k}$ can be generated by blowing up $\mathbb{F}_{2k-1}$
or $\mathbb{F}_{2k+1}$. Hence, to enumerate all possible base surfaces
with $r > 1$
that support elliptic Calabi-Yau threefolds 
we only need to take $\mathbb{F}_n$
with odd $n$ as starting points for the blowing up process.

The set of effective divisor classes on $S$ forms a cone, called
the {\em effective cone} Eff$(S)$. The dual of the effective cone is the set of
divisor classes that intersect non-negatively with $[C]\in$Eff$(S)$;
this is called the  {\em nef cone} Nef$(S)$:
\be
\mathrm{Nef}(S)=\{[D]\in\mathrm{Pic}(S)|
\forall [C]\in\mathrm{Eff}(S)\ ,\ [D]\cdot[C]\geq 0\}.
\ee

Elements of the nef cone are called nef divisors (classes). In
general, an element in the effective cone need not be nef; for
example, any effective divisor of negative self-intersection is not
nef.  Conversely, however (see {\it e.g.}  Corollary II.3 in
\cite{Harbourne1997}), any nef divisor class on
a smooth rational
surface is effective.

The effective cone is the primary distinguishing characteristic of
different rational surfaces.  For example, the Hirzebruch surfaces
$\F_n$ with $n$ odd all have the same intersection form as discussed
above.  The cone of effective divisors, however, is generated by
$[S_\infty]$ and $[F]$, and is different for each $n$.

Our primary tool in this paper for studying base surfaces $S$ will be
the combinatorial structure of the effective cone.  The following
result, which is Proposition 1.1 in \cite{Cox-rings},
will be useful
\begin{lemma}
\label{l:1}
For surfaces generated by blowing up $\mathbb{F}_n\,(n\leq 12)$, which
have Picard rank greater than 2, when the effective cone is polyhedral
({\it i.e.} generated by a finite set of vectors), then  the effective
cone is generated by rational divisor classes with negative
self-intersection.
\end{lemma}

For the Hirzebruch surfaces, of Picard rank 2, it follows directly
from the explicit description above that the effective cone is
generated by rational divisor classes with non-positive
self-intersection ({\it i.e.}, $[S_\infty], [F]$, where $[F] \cdot [F] = 0,
[S_\infty] \cdot [S_\infty] = -m$ for $\F_m$).  
After one blow-up,  Lemma~\ref{l:1} applies.
Blowing up $\F_m$ ($m$ odd),
an effective class
$[F]-[E_2]=[H]-[E_1]-[E_2]$ with $(-1)$ self-intersection always
appears. Then the effective class with 0 self-intersection can be
written as a
non-negative linear combination
of $[H]-[E_1]-[E_2]$ and
exceptional divisors. This property generally holds for surfaces with
higher Picard rank.

We discuss these statements further in section 4.

Note that curves of negative self-intersection on a base surface that
supports an elliptic Calabi-Yau must always be rational.  From
(\ref{RiemannRoch}), an irreducible curve of negative self-intersection with $g
\geq 1$ satisfies $- K \cdot C \leq C \cdot C < 0$.  From this it
follows that $- nK$ contains $C$ at least $n$ times.  This means that
$f, g$ must vanish to at least orders $4, 6$ over $C$ so we cannot
have such a curve in a base that supports an elliptic Calabi-Yau
threefold.

To close this section, we introduce some terminology that we will use
frequently in the remainder of the paper.
\begin{itemize}
\item{In general we will not distinguish between ``divisor class'',
  ``divisor'' and ``curves''.
Thus, for example, the divisor class
  $[D]$ is generally written as $D$.
When we talk about ``$n$-curves'', this refers to a
  divisor class with self-intersection $n$. 
We often use the term ``negative curve'' to refer to a curve of
negative self-intersection.
}

\item{Following   Lemma \ref{l:1} and the subsequent discussion, the
  negative curves that appear in the surfaces in which we
  are interested in are always rational.
Furthermore, in the cases where Lemma \ref{l:1} applies, we
  define Neg$(S)$ as the set of irreducible negative rational curves
  that generate Eff$(S)$. A general requirement for Neg$(S)$ is:

For $C,D\in\mathrm{Neg}(S)$, if $C\neq D$, then $C\cdot D\geq 0$. This
is just the statement that two different
irreducible curves cannot intersect each
other negatively.}

\item{We further define the set of irreducible rational curves $C$ with
  self intersection number less than -1 to be Sing$(S)$. 
These include the non-Higgsable clusters described in 
Table~\ref{t:clusters}, as well as configurations of $(-2)$-curves whose
  intersection matrices are exactly the Cartan matrices for ADE Lie
  algebras.}

\item{When discussing toric bases, we indicate the cycle of
  self-intersection numbers by {\it e.g.} $((n, 0, - n, 0))$.} 

\item{We will use the  vector representation
  $(a_0,a_1,\dots,a_{r})$ for the divisor class $D=a_0 H+\sum_{i=1}^r
  a_r E_r$. This applies to surfaces that arise from blowing up any
  $\F_{2k-1}$ any number of times.}

\item{We define the set of genus $g$, self-intersection $k\geq 0$
  divisors $C$
that intersect
  non-negatively with every curve $D\in\mathrm{Neg}(S)$ by
  $\mathfrak{C}_{g,k\geq 0}(S)\subset \mathrm{Nef}(S)$. They are all
  effective curves, as we mentioned before.
All such curves can be described as integer vectors as described in
the previous point.

Finally we define $\mathfrak{C}_k(S)=\bigcup_g \mathfrak{C}_{g,k}(S)$.}

\item{Using the terminology in \cite{Derenthal, DerenthalThesis}, for
  two surfaces $S$ and $R$, if there is a bijection between Neg$(S)$
  and Neg$(R)$ that preserves the intersection structure then
  Eff$(S)$ and Eff$(R)$ are isomorphic, and we say that $S$ and $R$ are
  ``of the same type''. This equivalence relation is weaker than
  isomorphism, as demonstrated in
  \cite{DerenthalThesis}. Nevertheless, two surfaces of the same type give
  the same minimal gauge algebra. The blow-up descendants of two
  surfaces of the same type may be different, as will be discussed in
  section 5.}
\item{We always denote the one-time blow-up of the surface $S$ by $S'$}
\end{itemize}

\section{The general strategy and some simple examples}
\label{sec:example}

Given the basic background and definitions from the previous section,
we can now outline the strategy we will use for constructing bases.
The idea is to keep track only of the combinatorial information
associated with the generators of the effective cone.
While this loses some information -- in particular it does not
distinguish surfaces of the same type that are not isomorphic -- it
gives us a simple combinatorial handle on a large class of surfaces.

Assume that we are given the information about the generators of the
effective cone for a given surface $S$. According to Lemma
\ref{l:1}, this will generally be a finite set of curves of
negative self-intersection.  We can then attempt to construct all
blow-ups of $S$ by considering all combinatorial possibilities
consistent with the geometric structure of the effective cone.  In
many cases this is straightforward.  The point $p$ that is blown up
can either lie on one of the negative curves that is a generator of
the effective cone, or on an intersection of such negative curves, or
on a more general point.  Lemma~\ref{l:passthrough} can be used to
determine which curves on the surface can pass through the point $p$,
and this information in principle can be used to determine the
structure of the effective cone for the blown up surface $S'$.

While this approach works in many situations in an unambiguous
fashion, there are various situations in which complications arise.
Such complications can include the appearance of an infinite number of
generators for the effective cone, or a situation as mentioned above
where there are non-isomorphic surfaces of the same type.  We discuss
these issues in more detail later in the paper.  Here we give a few
examples of how the combinatorial data describes the effective
cone and blow-ups in some  simple cases where there are no complications.
These examples serve as illustrations of the general ideas described
in the previous section, and clarify the nature of the computational
problem in
generalizing the construction to arbitrary surfaces.

Let us begin with the base $S =\P^2$.  The intersection form on this base
is positive definite, with generator $H$ having $H \cdot H = 1$.  The
divisor class of $H$, corresponding to a line on $\P^2$, is a generator
of the effective cone.
From (\ref{eq:h11}, \ref{eq:h21}), it is straightforward to determine
that the Hodge numbers of the generic elliptic fibration over $\P^2$
are $(2, 272)$, since $R = T = V= H_{\rm charged} =0$.

Now consider a single blow-up of $\P^2$ at a generic point $p$.  There is
a line on $\P^2$ that passes through  any point $p$ (a trivial
application of Lemma~\ref{l:passthrough}).  Thus, the generators of
the effective cone for the new surface $S'$ are
\begin{eqnarray}
[L]_{(0)}: &  &  (1, -1) \label{eq:f1}\\
{}[E]_{(-1)}: &  &  (0, 1)  \nonumber
\end{eqnarray}
where the subscript on the divisor class denotes the self-intersection
of that curve.  This is, of course, the Hirzebruch surface $\F_1$ as
described above, also known as the del Pezzo surface $dP_1$; in
general, the del Pezzo surface $dP_r$ is constructed by blowing up
$\P^2$ at $r$ independent points.  The base $\F_1$ has a toric description,
with a cyclic set of toric divisors having self-intersections $((+ 1, 0, -1,
0))$.
The generic elliptic fibration over this base has Hodge numbers $(3,
243)$, since $T = 1$.

Now, we consider possible blow-ups of the surface $S = \F_1$.
There are only two distinct types of blow-up, from the point of view
of the combinatorics of the effective cone.  We can blow up at a point
on the curve $E = (0, 1)$, or we can blow up at a generic point.    These two types of blow-ups result in the
following surfaces:

Blowing up $\F_1$ at a generic point $p_2$
that is not
on the $-1$ curve $E$ gives a new effective cone
generated by three $-1$ curves
\begin{eqnarray}
[L_{12}]_{(-1)}: & &  (1, -1, -1)\\
{}[E_1]_{(-1)}: & &  (0, 1, 0)\\
{}[E_2]_{(-1)}: & &  (0, 0, 1)
\end{eqnarray}
This surface is also known as $dP_2$.  Note that the line $L$ in
$\F_1$ has self-intersection 0, so by Lemma~\ref{l:passthrough} will pass
through any point, including the point $p_2$;
this guarantees the existence of the $-1$ curve $L_{12}$, which
corresponds in the original $\P^2$ to the line passing through the two
points $p_1, p_2$ that are blown up to get this surface.

Blowing up $\F_1$ at a point $q$
on the $-1$ curve $E$ gives a new effective cone
generated by two $-1$ curves and a $-2$ curve
\begin{eqnarray}
{}[L]_{(-1)}: & &  (1, -1, -1)\\
{}[E_1]_{(-2)}: & &  (0, 1, -1)\\
{}[E_2]_{(-1)}: & &  (0, 0, 1)
\end{eqnarray}
The fact that the line $L$ on $\F_1$ is in a linear system of
dimension 1, and can be chosen to pass through $q$, follows
again from Lemma~\ref{l:passthrough}.
We refer to the resulting surface as $\Sigma_2$.  

Note that the two distinct
blow-ups of $\F_1$ to $dP_2$ and $\Sigma_2$ correspond to the two
branches above $\F_1$ in Figure~\ref{f:Chain}.  Each of these two
surfaces has a toric description.
The fact that in each of these cases the set of generators of the
effective cone consists completely of curves of negative self
intersection is a consequence of Lemma~\ref{l:1}.
Both of these surfaces support generic elliptic Calabi-Yau threefolds
with Hodge numbers $(4, 214)$.

We can continue in this fashion.  There are four combinatorially
distinct ways to blow up the surface $dP_2$, corresponding
to blow-ups on $E_1, L_{12}, E_1 \cap L_{12}$, and a generic point.

Blowing up $dP_2$ at a generic point gives $dP_3$.  The effective cone for
$dP_3$ has six generators, corresponding to the 3 exceptional divisors
$E_1, E_2, E_3$ and the 3 lines $L_{12} = (1, -1, -1, 0), L_{23} = (1,
0, -1, -1), L_{13} = (1, -1, 0, -1)$.  This illustrates one of the
main challenges of doing this combinatorial blow-up construction
systematically: it is necessary at each step to identify all possible
new curves of negative self-intersection that can be produced at each
stage of the blow-up process.  Addressing this challenge is the
primary goal of the following section.

Three of the four ways in which $dP_2$ can be blown up have a toric
description; blowing up at $E_1, E_1 \cap L_{12}$, and a generic point
respectively give toric bases with a cyclic set of effective toric
divisors having self-intersections $((0, -1, -1, -2, -1, -1)), ((0, 0,
-2, -1, -2, -1)),$ and $((-1, -1, -1, -1, -1, -1))$.  (The last of these
is $dP_3$).  The final possibility, blowing up at a point in $L_{12}$
gives a non-toric base with the following generators for the effective
cone
\begin{eqnarray}
[L]_{-2}: &  &(1, -1, -1, -1) \\
{}[E_1]: &  &  (0, 1, 0, 0)\\
{}[E_2]: &  &  (0, 0, 1, 0)\\
{}[E_3]: &  &  (0, 0, 0, 1)
\end{eqnarray}

\begin{figure}
\centering
\includegraphics[height=7.5cm]{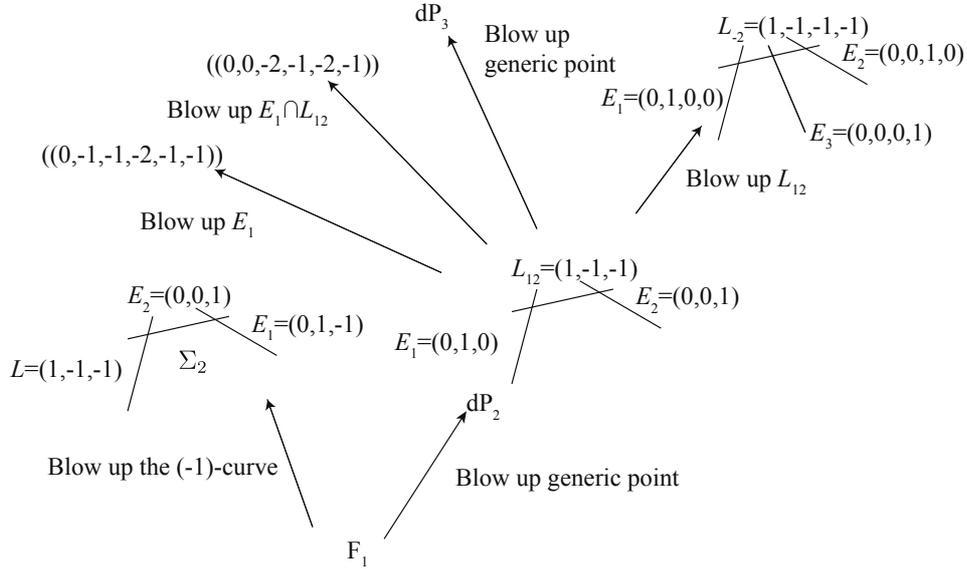}
\caption{The illustration of some of the blow up process discussed in this section, starting from $\mathbb{F}_1$.}\label{f:blowups}
\end{figure}

Similarly, there are 6 combinatorially distinct ways of blowing up the
surface $\Sigma_2$; 5 of these correspond to toric constructions, and
the sixth, associated with blowing up on the $-1$ curve $E_2$, gives
another non-toric base.
Of the toric bases, two involve blowing up a point on the $-2$ curve,
giving a $-3$ curve.  While the other bases support elliptic
Calabi-Yau threefolds with Hodge numbers $(5, 185)$, the two with $-3$
curves support elliptic Calabi-Yau threefolds with Hodge numbers $(7,
193)$, since there is a non-Higgsable cluster with $R = 2, V= 8$.

Continuing in this fashion, we can construct a wide range of types of
complex surfaces that support elliptic Calabi-Yau threefolds.  At each
stage it is necessary to determine the complete set of curves of
negative self-intersection that generate the effective cone.  Dealing
with this computation is the subject of the next section.

\section{Identifying new curves after a blow-up}
\label{sec:curves}

In this section we describe the different types of points that can be
blown up in passing from one surface $S$ to another $S'$.  We then
describe an algorithm to compute all  $-1$ curves that generate the
effective cone in finite time.

\subsection{Types of blowups}

As mentioned before, when we blow up a point  that lies 
with multiplicity $m$  on a
representative of an effective divisor class $C$ 
and self-intersection $k$, there is a new effective divisor class
$C'=(C,-m)$, written in the vector representation. Using the
Riemann-Roch formula (\ref{RiemannRoch}),  the following lemma
holds:
\begin{lemma}
If one blows up a curve $C\in\mathfrak{C}_{g,k}(S)$ at an $m$-point on
$S$, the resulting new effective curve on $S'$ is:
$C'=(C,-m)\in\mathfrak{C}_{g-(m^2-m)/2,k-m^2}(S')$.\label{l:blpC}
\end{lemma}

This implies that blowing up a single point on a curve will not change
its (arithmetic) genus, but blowing up an $m$-point on a curve will
decrease the genus of the curve by $(m^2-m)/2$.

Then given a  base $S$, the different ways of blowing it up can be
classified as follows:
\vspace*{0.1in}

\noindent
(1) One can blow up a generic point $p$ on $S$. Then all the elements
$C\in$ Neg$(S)$ are transformed to $C'=(C,0)$, because they are fixed
and they will not pass through a generic point $p$.
The exceptional curve $E' = (0, \ldots, 0, +1)$ is an element in
Neg$(S')$ that
is a generator of the new effective cone.
In order to construct
the full Neg$(S')$ that generates Eff$(S')$, we also need to blow up
all the elements $C\in\mathfrak{C}_{0,0}$ at a single point, blow up
all the elements $C\in\mathfrak{C}_{1,3}$ at a double point, blow up
all the elements $C\in\mathfrak{C}_{3,8}$ at a triple point, and so
on\ldots This process can always be done  in principle
due to  Lemma
\ref{l:passthrough}.
In the next subsection we describe how this can be realized in
practice with a finite amount of computation.

As a consistency requirement, the resulting rational curves after the
blow up should intersect non-negatively with each other. This is
equivalent to the following statement:

For $A\in\mathfrak{C}_{(m^2-m)/2,m^2-1}(S)$,
$B\in\mathfrak{C}_{(n^2-n)/2,n^2-1}(S)$ that are blown up in this
process, $A\cdot B\geq mn$ ($m,n>0$).
\vspace*{0.1in}

(2) One can choose to blow up a non-generic point, so that a set of
curves $C_{0,i}\in\mathfrak{C}_{0,k<0}$ are blown up at a single
point.
The index $i$ here just labels different curves
with same genus.  In the simplest cases, the point blown up lies only
on a single rational curve $C$ of negative self-intersection, in which
case the transformed curve is $ C' = (C, -1) $.  The point blown up
may also lie at an intersection point between a pair of negative
curves $C, D$, in which case both curves are transformed and $C' \cdot
D' = 0$.

(3) There are also
situations where a set of curves $C_{1,i}\in\mathfrak{C}_{1,k<3}$ are
blown up at a double point, $C_{2,i}\in\mathfrak{C}_{3,k<8}$ are blown
at a triple point, and so on.  This will produce new negative rational curves
of self-intersection $< -1$ that must be included in Neg$(S')$ We
define the blow-up process to be a ``special blow-up'' when one or
more (-2) or lower curves are generated by blowing up positive curves
at points with multiplicity higher than 1.

(4) In some cases it may be possible to choose a non-generic point
as in (2) that lies at the intersection of more than two negative
curves.
\vspace*{0.1in}

Cases (1) and (2) can be handled in a systematic fashion using the
combinatorial data of the effective cone. Case (3) may or may not give
us new bases. In the specific regimes that are thoroughly studied in
this paper, special
blow-ups are observed not to occur. 
Case (4) is discussed further in section
\ref{sec:obstructions}, and also does not occur in the regimes that we
study explicitly.

Just as for blowups producing $-1$ curves, we require that for
intersecting curves
$C_{(m^2-m)/2,i}\in\mathfrak{C}_{(m^2-m)/2,k<m^2-1}$,
$C_{(n^2-n)/2,j}\in\mathfrak{C}_{(n^2-n)/2,k<n^2-1}$, that are blown
up at points of multiplicity $m, n$ respectively, we must have
$C_{(m^2-m)/2,i}\cdot C_{(n^2-n)/2,j}\geq mn$. Otherwise after the blow up, Neg$(S')$ will contain two
different elements that intersect each other negatively.

\subsection{Generating the set of $(-1)$-curves on $S'$}

From the analysis above, all the curves in the set Sing$(S')$ can be
generated by blowing up non-generic points. However, there is also in
general a new set of (-1)-curves in Neg$(S')$, and it is not clear
whether the algorithm generating (-1)-curves described above is finite
or not. In principle we need to look at all the sets
$\mathfrak{C}_{(m^2-m)/2,m^2-1}$, but this is impossible to do. Now we
prove a powerful proposition, which contains the methodology to
generate all the $(-1)$-curves with a finite amount of work:
\begin{prop}
The set  $\mathfrak{C}_{0,-1}(S')$
of rational (-1)-curves $C=(a,b_1,b_2,\dots,b_r)\in \mathrm{Neg}(S')$
generated by the blow up method is the solution set to the following
Diophantine equations: 
\be
\bsp
a^2-\sum_{i=1}^r b_i^2\ =\ &-1\\
3a+\sum_{i=1}^r b_i\ =\ &1\label{Dioph1}
\end{split}
\ee
with the additional requirement that the curve intersects
non-negatively with all the elements in Sing$(S')$.\label{p:gcurves}
\end{prop}

Actually the two equations are nothing but the defining equations of
self-intersection and genus.

\begin{proof}
We prove this by induction.  First one can check the correctness of
this statement for $\mathbb{P}^2$ and $\mathbb{F}_n$. Then suppose
this is true for a base $S$; we want to show this is also true for
any surface $S'$ that is generated by blowing up $S$ once. 
\vspace*{0.1in}

\noindent
(i) All the curves $C'\in\mathfrak{C}_{0,-1}(S')$  satisfy the
requirements in the proposition.

Obviously they satisfy the Diophantine equations.  If $C'$
negatively intersects with $D'\in$ Sing$(S')$, this means $C'=D'+F'$
where $F'$ is some other effective divisor, hence $C'$ is reducible
and is not in Neg$(S')$.
\vspace*{0.1in}

\noindent
(ii) All the irreducible rational (-1)-curves $C'$ that satisfy the
non-negative intersection requirement in the proposition are elements
of $\mathfrak{C}_{0,-1}(S')$, and can be generated by the blow-up
process.  We analyze the different types of solution to (\ref{Dioph1})
separately.

(a) This is obviously true for the solution $(0,\dots,0,1)$ since it
is the exceptional curve $E'\in\mathfrak{C}_{0,-1}(S')$ associated
with the blowup $S \rightarrow S'$, and none of the other curves has a
positive last entry.

(b) For all the solutions of the form $C'=(C,0)$, 
where we know that $C$
intersects non-negatively with all the elements in Sing$(S)$, it follows by
induction that $C\in\mathfrak{C}_{0,-1}(S)$. From the general
characterization of the
blow-up process described earlier, $C'=(C,0)$ is still in Eff$(S)$, but it may
be represented as a positive linear combination of a lower
self-intersection effective curve and the exceptional curve:
$C'=(C,-m)+m (0,1)$.  When this happens, however,  it means that $C'$ intersects
negatively with $(C,-m)\in$ Sing$(S')$, which contradicts the
requirement in the proposition. Hence $C'=(C,0)$ is irreducible when
it satisfies the requirement in proposition, and it is in
$\mathfrak{C}_{0,-1}(S')$.

(c) For all the solutions of the form $C'=(C,-m)$, $m> 0$, $C$ is a
  genus $(m^2-m)/2$, self intersection $m^2-1$ divisor. In fact the
  solution to the equations (\ref{Dioph1}) automatically intersects
  non-negatively with any other solutions
(see Theorem 2a in \cite{Nagata}).  
Together with the assumption that $C'$ intersects non-negatively with
all the elements in Sing$(S')$, we can conclude that $C'$ intersects
non-negatively with any other curve in Neg$(S')$. From this we can
also know that $C$ is nef hence is effective on $S$. Then according to
Lemma \ref{l:passthrough}, $C\in\mathfrak{C}_{(m^2-m)/2,m^2-1}(S)$ can
be blown up at a generic point with multiplicity $m$. These statements
together guarantee that $C'$ is an irreducible curve on $S'$. Hence
any solution of the form $C'=(C,-m)$, $m> 0$ that satisfies the
requirement in the proposition is in Neg$(S')$.
\end{proof}

Actually, it is known that Proposition 1 holds for generalized del
Pezzo surfaces \cite{Derenthal, DerenthalThesis}. We just needed to confirm
that it holds for the  bases that we are interested in.

Now the question is how to generate the solutions to the Diophantine
equations using a finite algorithm. In fact, all the solutions can be
generated by a series of ``$q$-operations'' acting on curves, which are
defined as follows \cite{Nagata}:

For a curve in the form $(a,b_1,b_2,\dots,b_r)$, one picks three numbers
$i_1,i_2,i_3$ out of $\{1, \ldots, r\}$, then performs the following
transformation:
\be
a\rightarrow a+d\ ,\ b_{i_1}\rightarrow b_{i_1}-d\ ,\ b_{i_2}\rightarrow b_{i_2}-d\ ,\ b_{i_3}\rightarrow b_{i_3}-d\ ,\ d=a+b_{i_1}+b_{i_2}+b_{i_3}.\label{qoperation}
\ee
It can be explicitly checked that the two quantities $a^2-\sum_{i=1}^r
b_i^2$ and $3a+\sum_{i=1}^r b_i$ do not change under this
transformation. In fact, this operation is the Cremona transformation
with center $p_{i_1},p_{i_2},p_{i_3}$.

In practice, one starts from some low-degree curves, such as all the
degree 0 and degree 1 rational (-1)-curves. Then one  tries to perform
all the $q$-operations on one curve so that the degree is increased
($d>0$ in (\ref{qoperation}) ). If the curve intersects
negatively 
with some element in Sing$(S)$, then we cut this branch. This
recursive algorithm is finite if Eff$(S)$ is finitely generated. We
discuss more about the finiteness of curves in section 6.

There are some cases, however, where a high degree (-1)-curve exists
but there are no valid degree 1 curves.  If this happens, the set of
(-1)-curves we get from the recursive algorithm above is not complete.
This is equivalent to the statement that there is another (-1) curve
that intersects non-negatively with all the curves we have found, but
it is not in the solution set generated by carrying out the algorithm
up to a fixed degree $d$.  Such an additional curve would appear to be
a negative curve in the nef cone, which is impossible.  While the full
set of (-1)-curves would in principle be produced by keeping all
intermediate branches and
continuing the
algorithm to arbitrary $d$, without some further check on the
completeness of the set of generators for the effective cone we would
not have a means of terminating the algorithm.  So for a consistency
check we compute the dual cone of Neg$(S)$ generated by the method
above, and check that the generators (extremal rays) of this dual cone
are non-negative. If a generator is negative, we know it is a
(-1)-curve in Neg$(S)$ that was not generated by the $q$-process. After
adding all the (-1)-curves of this kind, we can get a complete set of
curves in Neg$(S)$, with a dual cone that contains no negative curves.

Another more complicated issue is the ``special blow-up'' that was
defined earlier, where one blows up a non-generic point on a curve
with genus higher than 1 to get an element in Sing$(S')$. This also
appears to make the algorithm infinite, since in principle we need to
examine the set of all positive curves.  This issue appears to be
difficult to handle in a completely general context, and we will
discuss this issue in specific regimes.

\section{Working example: (generalized) del Pezzo surfaces}
\label{sec:gdp}

As a simple set of examples we now consider the class of surfaces
called  del Pezzo surfaces and
generalized del Pezzo surfaces. The definition and basic
properties of these objects can be found in \cite{Derenthal}.

A generalized del Pezzo surface is a non-singular projective
surface whose anticanonical class $-K$ satisfies
$K \cdot K > 0$ and $-K \cdot D \geq 0$ for all effective divisors $D$
on the surface.

Any generalized del Pezzo surface is isomorphic to
$\mathbb{P}^2$, $\mathbb{P}^1\times\mathbb{P}^1, \F_2$, or the blow up of
$\mathbb{P}^2$ in $r\leq 8$ points in almost general position. The
degree of the generalized del Pezzo surface is given by $9-r$. The
phrase ``almost general position'' means that the surface does not
contain curves of self-intersection $-3$ or below, which implies that
it does not contain any NHCs and there is no minimal gauge content in
the corresponding low-energy 6D supergravity theory.

If a generalized del Pezzo surface does not contain any (-2)-curves,
then it is just an ordinary del Pezzo surface, which is
$\mathbb{P}^2$, $\mathbb{P}^1\times\mathbb{P}^1$, or the blow up of
$\mathbb{P}^2$ in $r\leq 8$ points in general position. By general
position, one requires that no three points in $p_1,\dots,p_r$ lie on
a line, no six points lie on a conic and no eight points lie on a
cubic with one of them a double point on that curve. This is
equivalent to the statement that none of the points $p_1, \ldots, p_r$
being blown up is a point on a (-1)-curve or a double point on a
2-curve, which would result in (-2)-curves. We define the del Pezzo
surface $dP_r$ to be the surface that comes from blowing up $r$ points
on $\mathbb{P}^2$.  This surface is also called Bl$_r \mathbb{P}^2$ or
$dP_{9-r}$ in different texts. Del Pezzo surfaces are the only
examples of 2D Fano varieties, which are surfaces with ample
anti-canonical class.
An
ample divisor is a positive divisor that intersects positively with
any irreducible curve on the surface, from the Nakai-Moishezon
criterion.

Generalized del Pezzo surfaces are the only \emph{almost Fano}
varieties among complex surfaces, where the anti-canonical class can
have vanishing but not negative intersection with irreducible curves.

In this section we focus on the structure of ordinary del Pezzo
surfaces; we return to a more detailed discussion of generalized del
Pezzo surfaces in \S\ref{sec:small}.
The set Neg$(S)$
for the ordinary del Pezzo surfaces $dP_r$ is well
understood; see \cite{Manin} for example. This corresponds for each $r$ to the
solution set to the Diophantine equations (\ref{Dioph1}), without
imposing any further constraints.

When $r=8$, the recursive algorithm using $q$-operations is finite, and
gives
only the following types of solutions:
\be
\bsp
&(0,1,0,0,0,0,0,0,0)\\
&(1,-1,-1,0,0,0,0,0,0)\\
&(2,-1,-1,-1,-1,-1,0,0,0)\\
&(3,-2,-1,-1,-1,-1,-1,-1,0)\\
&(4,-2,-2,-2,-1,-1,-1,-1,-1)\\
&(5,-2,-2,-2,-2,-2,-2,-1,-1)\\
&(6,-3,-2,-2,-2,-2,-2,-2,-2)\label{dP8curves}
\end{split}
\ee
up to permutations on the entries $a_1, \ldots, a_r$. For smaller $r$,
the possible solutions are the truncations of these vectors by
deleting some ``0'' entries.  For $r\geq 9$, the recursive algorithm
is infinite, as there are an infinite number of types of solutions on
$dP_9$.  We list the number of (-1)-curves on del Pezzo surfaces with
each value of $r \leq 9$ in Table \ref{t:dPcurves}.  Note that when
the 9th point blown up is chosen so that it lies at the intersection
of two cubics that pass through the first 8 points, we get a special
class of surface known as a rational elliptic surface.  In these
cases, $dP_9$ can act as a good base for a CY threefold, though the
effective cone is not finitely generated.  We discuss some aspects of
this in the following sections.  When the 9th point blown up is not
the special common point to the cubics passing through the first 8,
then $-K = (3, -1, \ldots, -1)$ corresponds to a rigid irreducible
genus 0 curve.  It follows that for an F-theory construction $f, g$
would vanish to orders $4, 6$ on this curve so this could not be a
good F-theory background.  For the same reason, a good base cannot be
formed after more than nine blow-ups without any non-Higgsable
clusters.
\begin{table}
\begin{center}
\begin{tabular}{|c|c|c|c|c|c|c|c|c|c|}
\hline
r&1&2&3&4&5&6&7&8&9\\
\hline
$|\mathrm{Neg}(dP_r)|$&1&3&6&10&16&27&56&240&$\infty$\\
\hline
\end{tabular}
\end{center}
\caption[x]{\footnotesize The number of (-1)-curves on each of the del
  Pezzo surface $dP_r$. There are no curves of lower self-intersection
  on $dP_r$, hence these numbers are equal to
  $|\mathrm{Neg}(dP_r)|$. We generalize the notation of $dP_r$ to
  include $r=9$, representing the surface that arises from blowing up
  9 general points on $\mathbb{P}^2$.}\label{t:dPcurves}
\end{table}

It is an instructive exercise to generate the negative curves by the
methodology of blow-ups described in the last section, with ordinary del
Pezzo surfaces as example. At each step we explicitly describe the set
of curves whose blow-ups give new $-1$ curves.
Suppose we start from $dP_2$, with
\be
\bsp
&\mathrm{Neg}(dP_2)=\{(0,1,0),(0,0,1),(1,-1,-1)\},  \; \mathfrak{C}_{0,0}(dP_2)=\{(1,-1,0),(1,0,-1)\},\\
&\mathfrak{C}_{1,3}(dP_2)=\mathfrak{C}_{3,8}(dP_2)=\dots=\varnothing \,.
\end{split}
\ee
To get $dP_3$, we blow up a generic point of $dP_2$. Then the generators
of the effective cone for $dP_3$ are of the following three types:
the exceptional curve $E$; $(C,0)$ with $C\in\mathrm{Neg}(dP_2)$, and
$(D,-1)$
with $D\in\mathfrak{C}_{0,0}$.  Together these give:
\be
\bsp
\mathrm{Neg}(dP_3)&=\{(0,1,0,0),(0,0,1,0),(1,-1,-1,0),(1,-1,0,-1),(1,0,-1,-1),\\
&(0,0,0,1)\}.
\end{split}
\ee
Then with
\be
\bsp
&\mathfrak{C}_{0,0}(dP_3)=\{(1,-1,0,0),(1,0,-1,0),(1,0,0,-1)\}, \; \\
&\mathfrak{C}_{1,3}(dP_3)=\mathfrak{C}_{3,8}(dP_3)=\dots=\varnothing,
\end{split}
\ee
we can similarly generate all the elements in $\mathrm{Neg}(dP_4)$:
\be
\bsp
\mathrm{Neg}(dP_4)\ =\ &\{ (0,1,0,0,0),(0,0,1,0,0),(0,0,0,1,0),(1,-1,-1,0,0),\\
&(1,-1,0,-1,0),(1,0,-1,-1,0),(1,-1,0,0,-1),\\
&(1,0,-1,0,-1),(1,0,0,-1,-1),(0,0,0,0,1)\} .
\end{split}
\ee
Furthermore
\be
\bsp
&\mathfrak{C}_{0,0}(dP_4)=\{(1,-1,0,0,0),(1,0,-1,0,0),(1,0,0,-1,0),(1,0,0,0,-1),\\
&(2,-1,-1,-1,-1)\},\ \mathfrak{C}_{1,3}(dP_4)=\mathfrak{C}_{3,8}(dP_4)=\dots=\varnothing,
\end{split}
\ee

The number of (-1)-curves on $dP_5$ hence is the number of (-1)-curves
on $dP_4$ plus the number of elements in $\mathfrak{C}_{0,0}(dP_4)$
plus 1, which is 16. This set perfectly reproduces the list of (-1)-curves
that come from solving the Diophantine equation.

$\mathfrak{C}_{0,0}(dP_5)$ consists of curves of the form $H-E_i (1\leq
i\leq 5)$ and $2H-E_i-E_j-E_k-E_l(1\leq i<j<k<l\leq 5)$. Hence
there are in total 10 curves in this set.

The number of (-1)-curves on $dP_6$ hence is $16+10+1=27$, which
exactly corresponds to the 27 lines on cubic surface.

Actually, $dP_6$ is the first case with a non-empty $\mathfrak{C}_{1,3}$;
there is exactly one element in this set: $(3,-1,-1,-1,-1,-1,-1)$, and the
number in $\mathfrak{C}_{0,0}(dP_6)$ is 27. Hence the number of
(-1)-curves in $dP_7$ is the number of curves in
$\mathfrak{C}_{0,0}(dP_6)$ plus the number of (-1) curves on $dP_6$
plus 1 in $\mathfrak{C}_{1,3}(dP_6)$ plus the exceptional curve, which
is $27+27+1+1=56$.

$dP_7$ is the first case with non-empty $\mathfrak{C}_{3,8}$; there is
exactly one element in this set:
$$(6,-2,-2,-2,-2,-2,-2,-2).$$
The number of (-1) curves on $dP_8$ is given by
\be
\bsp
|\mathrm{Neg}(dP_8)|\ =\ &|\mathrm{Neg}(dP_7)|+|\mathfrak{C}_{0,0}(dP_7)|+|\mathfrak{C}_{1,3}(dP_7)|+|\mathfrak{C}_{3,8}(dP_7)|+1\\
\ =\ &56+126+56+1+1=240
\end{split}
\ee

\section{Geometric obstructions}
\label{sec:obstructions}

As mentioned in section 2, there is a subtlety that arises in some
situations, where the data on the intersection matrix of curves on
$S$ does not completely determine the possible ways to blow up the
surface.

\begin{figure}
\centering \includegraphics[height=7cm]{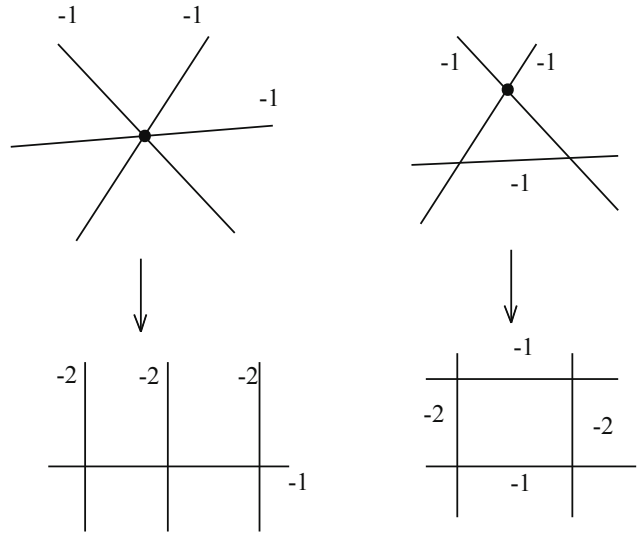}
\caption{Two geometric configurations with the same intersection structure
  and vector representations of curves, but which can be blown up to
  give surfaces with distinct intersection
  structure}\label{f:ambiguity}
\end{figure}

Suppose we have 3 negative curves that each intersect one another
pairwise.  Then, for example, we cannot distinguish the two cases
depicted in Figure \ref{f:ambiguity} given only the vector
representations of the curves and their corresponding intersection
matrix.  In one case there are three (-1)-curves that all intersect at
a single common point, and in the other case there are three distinct
intersection points where each pair of curves intersect.  But the blow
up procedures for these two configurations are different.  The case in
which three negative curves intersect at a single point can in many
situations be regarded as a special point in the moduli space of the
surface (\emph{e.g.}  dP$_6$) defined by Eff$(S)$. In principle, this
piece of information needs to be added along with the vector
representation of the curves in Eff$(S)$ in order to fully specify the
base, and this characterization of bases is finer than the notion that
the surfaces are ``of the same type''. In most situations in which
configurations of this type occur, one can choose whether the case on
the left or the case on the right in Figure \ref{f:ambiguity}
describes the geometry. In other cases of this type, however, it can
occur that only one of the geometric possibilities can actually be
realized.  Distinguishing which of the possibilities is geometrically
allowed seems in general to be a difficult problem.  For the explicit
classes of bases that we enumerate later in this paper, this kind of
situation does not arise.  But for a completely general analysis of
bases giving elliptic Calabi-Yau threefolds with arbitrary Hodge
numbers, a systematic methodology is needed for dealing with 
configurations with these kinds of issues.  We give some simple
examples here to illustrate the kinds of problems that can arise.

\begin{table}
\centering
\begin{tabular}{|c|cccccccccc|}
\hline
curve&point&a&b&c&d&e&f&g&h&i\\
\hline
A&1&-1&-1&-1&0&0&0&0&0&0\\
B&1&0&0&0&-1&-1&-1&0&0&0\\
C&1&0&0&0&0&0&0&-1&-1&-1\\
\hline
D&1&-1&0&0&-1&0&0&-1&0&0\\
E&1&0&-1&0&0&-1&0&0&-1&0\\
F&1&0&0&-1&0&0&-1&0&0&-1\\
\hline
G&1&-1&0&0&0&0&-1&0&-1&0\\
H&1&0&-1&0&-1&0&0&0&0&-1\\
I&1&0&0&-1&0&-1&0&-1&0&0\\
\hline
\end{tabular}
\caption[x]{\footnotesize A configuration of Sing$(S)$, with 9
  $(-2)$-curves that forms 3 groups of 3 curves that
intersect each
  other}\label{t:3I3}
\end{table}

\begin{figure}
\centering
\includegraphics[height=5cm]{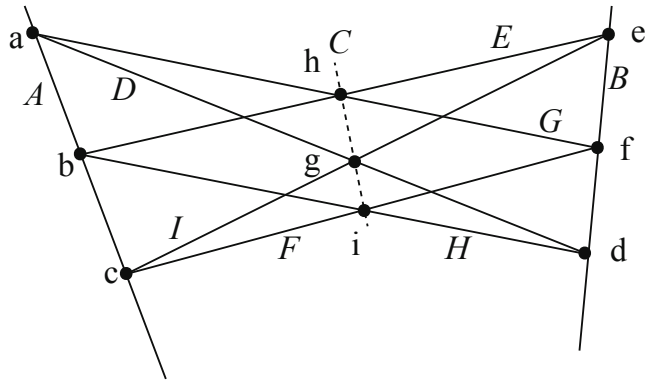}
\caption{The geometry of the blow-up points $a$,$b$,$c$,$d$,$e$,$f$,$g$,$h$,$i$, of the configuration in Table \ref{t:3I3}.}\label{f:3I3}
\end{figure}

One particular situation of this type that can arise involves
combinations of three negative curves that have to intersect at a
single point. For example, consider the configuration of Sing$(S)$,
with $r=9$, in Table \ref{t:3I3}; the corresponding geometrical
picture is drawn in Figure \ref{f:3I3}.  Note that in the last blow-up
at point $i$, lines $gh$, $cf$ and $bd$ are guaranteed to intersect at
a single point, by \emph{Pappus's theorem}.

This phenomenon is a special case of the more general
\emph{Cayley-Bacharach theorem} (for a complete overview of this
subject, see \cite{Eisenbud1996}).  Theorem CB4 in \cite{Eisenbud1996}
says that if 2 curves $A$ and $B$ of degree $d$ and $e$ intersect at
$d\cdot e$ distinct points $\Gamma=\{p_1,\dots,p_{de}\}$, and another
curve $C$ of degree $f \leq$  $d+e-3$ passes through
$d\cdot e-1$ points in $\Gamma$, then $C$ passes through all points in
$\Gamma$.

The \emph{Charles theorem} corresponds to the case $d=e=f=3$, which in
general states that if two cubic curves intersect at 9 points, then
another cubic curve that passes through 8 of these 9 points must pass
through the ninth point.  Pappus's theorem corresponds to the
degenerate case where these three cubic curves are linear combinations
of 3 lines respectively: $A+B+C$, $D+E+F$ and $G+H+I$ as in Table
\ref{t:3I3}.  Note that the more general case of Charles's theorem is
operational in the construction of a rational elliptic surface, where
after blowing up eight general points on $\P^2$, the ninth point blown
up is the common point to a pencil (one-parameter family) of cubics
that pass through the first eight points.

\begin{table}
\centering
\begin{tabular}{|c|ccccccccc|}
\hline
curve&point&a&b&c&d&e&f&g&h\\
\hline
A&1&-1&-1&-1&0&0&0&0&0\\
B&1&0&0&0&-1&-1&-1&0&0\\
C&1&-1&0&0&-1&0&0&-1&0\\
D&1&0&0&-1&0&-1&0&-1&0\\
E&1&0&-1&0&0&0&-1&-1&0\\
F&1&0&-1&0&0&-1&0&0&-1\\
G&1&-1&0&0&0&0&-1&0&-1\\
H&1&0&0&-1&-1&0&0&0&-1\\
\hline
\end{tabular}
\caption[x]{\footnotesize A configuration of Sing$(S)$, with 8 $(-2)$-curves}\label{t:4I3}
\end{table}

\begin{figure}
\centering
\includegraphics[height=5cm]{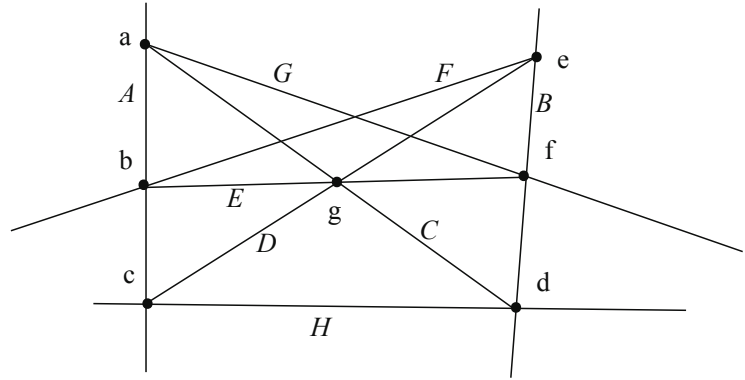}
\caption{The geometric picture of blowing-up $a$,$b$,$c$,$d$,$e$,$f$,$g$ of the configuration in Table \ref{t:4I3}.}\label{f:4I3}
\end{figure}

On the other hand, sometimes it is impossible for three negative
curves to intersect at a single point. Consider the configuration with
$r=8$ in Table \ref{t:4I3}.  We draw the first 7 blow up points on a
plane in Figure \ref{f:4I3}.  We then blow up the point $g$ with the
intersection properties described in Table \ref{t:4I3}. This requires
that $ad$, $ce$ and $bf$ all intersect at a single point.  From the
combinatorics of the effective cone, it seems possible to blow up an
eighth point $h$ to get curves $F,G,H$ with the intersection
properties described in Table \ref{t:4I3}.  In this situation,
however, there is requirement that $af$, $be$ and $cd$ intersect at a
single point, which is geometrically impossible. Hence the
configuration in Table \ref{t:4I3} is not allowed.
Thus, the structure of the combinatorics of the effective cone is in
general insufficient to guarantee that a set of three curves either
can or cannot all intersect at a single point.

For a complete analysis of all bases, we would need a systematic way
to distinguish between the following three possibilities
in any given situation:

(1) Three negative curves $A,B,C$  must all intersect at a single point.

(2) Three negative curves $A,B,C$ cannot  all intersect at a single point.

(3) Three negative curves $A,B,C$ may or may not intersect at a single point.

It seems possible that a systematic analysis using the generalized
Caylay-Bacharach theorem can give a clear general answer in these
situations, but we leave this to future studies.

One interesting aspect of these geometric issues is the question of
how the allowed combinatorial structures for the effective cone can be
understood from the point of view of the low-energy 6D supergravity
theory.  We return to this issue in  the final section.

\section{Finiteness of curves}
\label{sec:finite}

Another difficulty that can arise in the complete classification of
bases is the appearance of bases in which the cone of effective curves
Eff$(S)$ has an infinite number of generators.  In our classification
program, in particular, the finite generation property of Eff$(S)$ is
a prerequisite in Lemma \ref{l:1}.  The finiteness of (-1)-curves on
any given surface is, however, considered to be an open problem. A nice
partial result in this direction is the theorem proved in
\cite{Testa}, which states that any surface with big anticanonical
class has finitely generated Cox ring, which implies the finite generation
of Eff$(S)$ and finiteness of (-1)-curves (for more information about
Cox ring and finiteness see \cite{Ottem}). In the context of rational
surfaces with effective anticanonical class, the phrase ``big'' in this
context actually implies that one of the following conditions holds:

(1) $r\leq 8$

(2) $-K_S=A+B$, where $A=H-\sum_{i\in S_A} E_i$ and $B=2H-\sum_{i\in S_B}
E_i$ are two effective curves in Eff$(S)$, and either $ab=0$ or
$\fracs{1}{a}+\fracs{4}{b}>1$. Here the the two sets of indices satisfy $S_A\bigsqcup S_B=\{1,2,\dots,r\}$, where  ``$\bigsqcup$'' means disjoint union.

(3) $-K_S=A+B+C$, where $A=H-\sum_{i\in S_A} E_i$, $B=H-\sum_{i\in S_B} E_i$
and $A=H-\sum_{i\in S_C} E_i$ are three effective curves in Eff$(S)$, and
either $abc=0$ or $\fracs{1}{a}+\fracs{1}{b}+\fracs{1}{c}>1$. Here the three sets of indices satisfy $S_A\bigsqcup S_B\bigsqcup S_C=\{1,2,\dots,r\}$.

From the perspective of our recursive algorithm of finding all the
(-1)-curves, these conditions can be understood in the following way:

If the condition (1) holds, then on average the entries $b_i$ in the
$q$-operation increase by $\fracs{3}{r}>\fracs{1}{3}$ after a
$q$-operation, which increases the degree of the curve by one. If one
applies the $q$-operation sufficiently many times, then the sum of the
negative of any three entries $b_i$ will exceed $a$, and then there is
no $q$-operation increasing the degree of curve. The sufficiency of the
conditions (2) and (3) for guaranteeing a finitely generated effective
cone can be argued in a similar way.  For case (2) the way to
perform a $q$-operation while preserving the intersection number with $A$
and $B$ is to choose $i_1\in S_A$, $i_2,i_3\in S_B$. Then
on average the entries $b_i$($i\in S_A$) increase by $\fracs{1}{a}$,
and the entries $b_i$($i\in S_B$) increase by $\fracs{2}{b}$. Because
$\fracs{1}{a}+\fracs{4}{b}>1$, the  set of
$q$-operations that increase the
degree of the curve will not last forever. For case (3), the way  to perform a $q$-operation while preserving the intersection number with $A$,
$B$ and $C$ is to choose $i_1\in S_A$, $i_2\in S_B$, $i_3\in S_C$. On
average the entries $b_i$($i\in S_A$) increase by $\fracs{1}{a}$, the
entries $b_i$($i\in S_B$) increase by $\fracs{1}{b}$, and the entries
$b_i$($i\in S_C$) increase by $\fracs{1}{c}$. The possibility of
having an arbitrarily high degree curve is also excluded by the condition
$\fracs{1}{a}+\fracs{1}{b}+\fracs{1}{c}>1$.  

Although most bases that we are interested in for F-theory
constructions have finitely generated effective cones, one can also
construct surfaces with an infinite number of (-1)-curves, which may
nonetheless be good bases for F-theory compactification. For example,
we mention above the case of a rational elliptic surface.  An explicit
example is
an $r=9$  base $S$ with
\begin{eqnarray}
\mathrm{Sing}(S)
 & = & \left\{(1,-1,-1,-1,0,0,0,0,0,0),(1,0,0,0,-1,-1,-1,0,0,0),
\right.\nonumber\\
& &\hspace*{0.1in}\left.(1,0,0,0,0,0,0,-1,-1,-1)\right\} \,.
\end{eqnarray}
This is a generalized del Pezzo surface with three -2 curves. This surface
clearly satisfies the condition that $-K$ is in the effective cone,
but also admits an infinite number of $(-1)$-curves on the boundary
of the effective cone. We can prove this as follows: suppose we pick
an exceptional curve $C_0=(0,1,0,0,0,0,0,0,0,0)$. Then at step 1 we
perform the $q$-operation with entries $b_1,b_4,b_7$. At step 2 we
perform the
$q$-operation with entries $b_2,b_5,b_8$. At step 3 we perform
$q$-operation with entries $b_3,b_6,b_9$, and we infinitely repeat
these three steps. The intersection product with the three curves in
Sing$(S)$ is invariant during this process. The degree of the curve at
step $2n-1$ can be explicitly computed, and is equal to $n^2$. This
means that the process will give arbitrarily high degree curves, hence
the number of (-1)-curves is infinite.

Bases with an infinite number of generators for the effective cone are
difficult to incorporate into the kind of analysis we are doing here.
We cannot compute the dual cone of Neg$(S)$, nor study the blow
up of the surface in an efficient way. It would be
possible in practice to avoid this problem by stopping when the number
of curves produced by the algorithm exceeds a certain number, assuming
that in this case the surface contains an infinite number of
(-1)-curves, and the branch can be discarded.  It seems based on some
limited evidence that the only situations where an infinite number of
negative self-intersection curves arise is on a relatively small
number of limiting bases that cannot be further blown up, and which
are associated with Calabi-Yau threefolds having quite small $h^{2,
  1}$.  And it is not clear whether bases with this infinite number of
generators really give acceptable F-theory models.  We leave a further
exploration of these questions, however, to future work.

\section{The algorithm}
\label{sec:algorithm}

In this section we give  a finite recursive algorithm for generating
general bases,  with specific prescriptions for dealing with the
various potential complications discussed in preceding sections.

(1) We start from a base $S$, with Picard rank $r+1$ and a vector
representation of negative curves Neg$(S)$, Sing$(S)\in$ Neg$(S)$. This
data should always be finite.

(2) We construct all blowups of $S$ in three possible ways: blow up
the intersection point of two curves $C_i,C_j\in $ Neg$(S)$, blow up a
generic point on a curve $C_i\in$ Neg$(S)$, or blow up a generic point
on the plane.  We do not consider blowups at points at which three or
more negative curves intersect. The special blow-ups are also excluded
in the current algorithm.

After this step we have the new Sing$(S')$.

(3) With this new Sing$(S')$, we use the $q$-process to generate the
(-1)-curves on $S'$. In practice, we just start from all the curves of
degree 0 and degree 1. If the number of curves in this step reaches a
certain large value (such as 500 or 1,000), then the number of generators is considered to be
infinite and the base is discarded. This situation 
turns out not to occur in the 
regimes of this paper.

(4) Then we need to check if the degrees of vanishing of $f\in O(-4K),
g\in O(-6K), \Delta\in O(-12K)$ are greater or equal than $(4,6,12)$
on any of the divisors (see Table \ref{t:Kodaira}). If this happens
then this singularity cannot be resolved and the base is discarded.

Practically, we use the method of Zariski decomposition, as in
\cite{Classbasis}. The actual algorithm is, we decompose $-nK$ to
\be
-nK=\sum_{i=1}^k a_i C_i+Y.
\ee
The integral coefficients $a_i$ indicate the orders of vanishing of
$-nK$ on the divisors $C_i$. And $C_i(1\leq i\leq k)$ are all the elements in
Neg$(S')$. The residual part $Y$ should be an effective
$\mathbb{Q}$-divisor, which intersects non-negatively with all curves
$C_i$. We start from $a_1=a_2=\dots=a_k=0$, and examine $Y\cdot C_i$
for every $C_i$. If this quantity is negative for $C_i$, then we add a
minimal value to $a_i$ that will make this quantity non-negative and
do the check again, until $Y\cdot C_i\geq 0$ for every $C_i$. If in
the process any $a_i$ reaches a certain value (11 for $n=12$), then
the singularity is too bad to be resolved. When this happens the
process stops and the base is discarded.

If there is set of coefficients $a_i$ that pass the check, then
we examine all the intersection points of pairs of negative curves $a_i$ and
$a_j$. If the sum of coefficients $a_i+a_j>10$ for $n=12$, then this
intersection point needs to be blown up.

Furthermore, bases containing (-9),(-10),(-11)-curves are not good,
since there are always $(4, 6)$ points on these curves.  Such points
need to be blown up until the curve of large negative
self-intersection becomes a (-12)-curve.  (Note that the points blown
up in this process could be generic points on these curves or points
where they intersect with other negative curves). The base is good
only when no more points of this kind need to be blown up.

Applying this method, one can derive the restrictions on curves and
intersecting curve configurations to give all the connected subgraphs
that can appear without any (-1)-curves. These are just the NHCs
listed in Table \ref{t:clusters}, along with configurations with only
(-2)-curves. The latter case corresponds to Dynkin diagrams of $ADE$
Lie algebras and affine $ADE$ Lie algebras.  There are additional
restrictions on the ways in which different NHCs can consistently
connect to one other. The NHC linking rules listed in
\cite{Classbasis}, augmented by a further set of branching conditions
on curves of self-intersection below $-1$ in \cite{semi-toric}, give a
complete set of such rules.  In our analysis here, these rules
automatically are enforced by the conditions that there are no $(4,
6)$ curves or points in the base.

(5) If no additional points need to be blown up, then  we compute the
generators of the dual cone of Neg$(S')$. This is known to be a hard
problem, the exact algorithm is described in \cite[p.~11]{Fulton}. If
the vectors in Neg$(S')$ are $d=r+1$ dimensional, then one computes the
normal vector $u$ to each of the $(d-1)$-dimensional facets. Then if
$u$ or $-u$ intersects non-negatively with all $C\in\mathrm{Neg}(S')$,
then $u$ or $-u$ is a generator of Nef$(S')$. Hence if $n= | $Neg$(S')|$,
then the computational complexity is at least
$\binom{n}{d-1}=\binom{n}{r}$. This turns out to be the major
computational difficulty in this program.

After all the generators of the dual cone of Neg$(S')$ are found,  we check if
all the generators are non-negative. If not, we
add the negative ones into
Neg$(S')$. We repeat the step (4)(5), and finally the dual cone of Neg$(S')$ will be free of negative generators, which means that the set Neg$(S')$ contains the complete set of (-1)-curves.

We then further check if $-K=(3,-1,-1,\dots,-1)$ intersects
non-negatively with all the generators. If not, then $-K$ is not in
the effective cone hence the base is not allowed.

(6) If the previous test is passed, then this base $S'$ is good. The
next step is checking if the intersection structure is isomorphic to
one of the bases generated before. The graph isomorphism problem is
also known to be hard; it is not clear if there exists a polynomial
algorithm. In practice we used the ``VFlib'' library developed by
Pasquale Foggia \cite{vflib}.

(7) If all the tests are passed, add this base to solution set and restart step (1) using $S'$.

The overall starting points are the bases with $r = 2$ that come from blowing up
$\mathbb{F}_n$,  and which have no (-9),(-10),(-11) curves. We
list their Neg$(S)$ below (the l.h.s. is the cyclic toric diagram
for these bases $S$):
\be
\bsp
((0,0,-1,-1,-1)):&\{(0,1,0),(0,0,1),(1,-1,-1)\}\\
((1,-1,-1,-2,0)):&\{(0,1,-1),(0,0,1),(1,-1,-1)\}\\
((2,-1,-1,-3,0)):&\{(-1,2,0),(0,0,1),(1,-1,-1)\}\\
((3,-1,-1,-4,0)):&\{(-1,2,-1),(0,0,1),(1,-1,-1)\}\\
((4,-1,-1,-5,0)):&\{(-2,3,0),(0,0,1),(1,-1,-1)\}\\
((5,-1,-1,-6,0)):&\{(-2,3,-1),(0,0,1),(1,-1,-1)\}\\
((6,-1,-1,-7,0)):&\{(-3,4,0),(0,0,1),(1,-1,-1)\}\\
((7,-1,-1,-8,0)):&\{(-3,4,-1),(0,0,1),(1,-1,-1)\}\\
((11,-1,-1,-12,0)):&\{(-5,6,-1),(0,0,1),(1,-1,-1)\}\\
\end{split}
\ee

So $\mathbb{P}^2$ and $\mathbb{F}_n$ are not counted in the solution
set and they have to added in by hand.
\vspace*{0.1in}

Using this algorithm can produce a finite list of possible bases
in a specific desired range.
The possible sources of incompleteness or inaccuracy in such a list
are as follows:

(i) Bases where the effective cone has an infinite number of
generators, or a very large number that exceeds the arbitrary cutoff
in the code, will be missed.

(ii) Bases that are produced by  ``special blowups'' giving curves of
self-intersection -2 or below, or by blowing up points at the
intersection of more than two negative curves, will be missed by this
algorithm.

(iii) If there are bases in which, as in Pappus's theorem, a certain
combination of more than two negative self-intersection curves are
forced to intersect at a common point, this algorithm may generate
spurious bases that appear to have an acceptable combinatorial
structure for the cone of effective curves, but which do not
correspond to actual surfaces.

In the following sections we present some classification results of
bases under certain limits. We have checked that in these regimes none of
the subtleties just mentioned occurs, so we indeed generate all the
possible bases in these regimes.

\section{Classification of all bases with rk(Pic$(S))<8$}
\label{sec:small}

The first subset that we enumerate explicitly is the set of bases with
rk(Pic$(S))= h^{1, 1} (S) = T +1<8$.  We wish to explicitly determine
the set of all combinatorially distinct effective cones that are
possible for bases in this range; this determines all topological
types of bases, for each of which we can compute Hodge numbers for the
generic elliptic fibration over that base.
For this set of bases, none of
the problematic issues are relevant.  The smallest value of
rk(Pic$(S))$ where the issue arises that three negative curves may
intersect each other is 7, where we can have $A=(1,-1,-1,0,0,0,0)$,
$B=(1,0,0,-1,-1,0,0)$, $C=(1,0,0,0,0,-1,-1)$. Hence this subtlety
will not affect the classification of types of bases with
rk(Pic$(S))<8$.
Note, however, that at rk(Pic$(S))= 7$, a situation arises already
where there are non-isomorphic surfaces of the same type.  On $\P^2$,
we can blow up six points in such a way that we can realize either of
the two configurations in Figure \ref{f:ambiguity}.  These surfaces
are treated identically in our combinatorial analysis.  Blowing up
these non-isomorphic surfaces gives topologically different surfaces
with rk(Pic$(S))= 8$, so a systematic way of treating this issue would
be necessary to continue this analysis to higher values of $h^{1, 1}
(S)$. 
Within this range of consideration, the problem of special blow-ups
also is not a problem, since the shortest vector representation of a curve that
comes from a special blow-up is $(3,-1,-1,-1,-1,-1,-1,-1,-2)$.  And
the number of generators of the effective cone is finite for all
surfaces in this range.  Thus, the
list of all bases with
rk(Pic$(S))<8$  generated by our algorithm is indeed complete.

\begin{figure}
\begin{center}
\includegraphics[width=10cm]{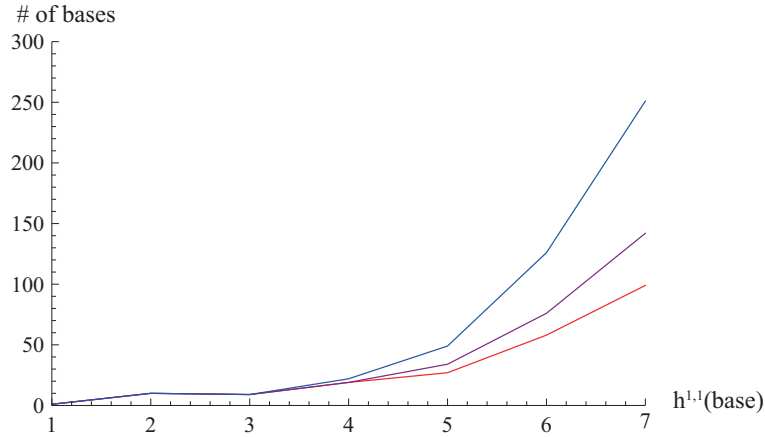}
\end{center}
\caption[x]{\footnotesize The number of bases with each value of
  $h^{1, 1} (S) < 8$
that are toric (red), semi-toric (purple), and
general including non-toric (blue).  }
\label{f:low}
\end{figure}

In total there are 468 bases with $8 >$ rk(Pic$(S))>0$. This includes
the surfaces $\mathbb{P}^2$ and $\mathbb{F}_n$ with $n=0\sim 8$ or 12.
Among these, 245 are non-toric and 177 are not included in the
semi-toric list in \cite{semi-toric}.
All previously identified toric and semi-toric bases appear in the set
generated by our algorithm, which gives a check on the  correctness
and completeness of
the result.
\footnote{Note that  for the purposes of counting and comparison in
  this paper
we include among toric and semi-toric bases those bases that have
toric or semi-toric structure including curves of self-intersection
$-9, -10, -11$; in some cases  such bases are not  really
toric or semi-toric
once the (4, 6) singularities on the $-9$ etc. curves are blown up.
We include these bases however in the toric and semi-toric sets since
they can be generated effectively using the corresponding toric or
semi-toric approach.
This technical distinction is discussed further in \cite{semi-toric}.}
A detailed breakdown of the number of bases with each value of Picard
rank that fit in each category is given in Figure~\ref{f:low}.

At low values of $h^{1, 1} (S) =r+1$, the set of bases generated can
be understood directly along the lines of the discussion in Section
\ref{sec:example} and Figure~\ref{f:blowups}.  For $r =0, 1$ the only
bases are $\P^2$ and the Hirzebruch surfaces.  The blowups to $r = 2$
give a set of 9 toric surfaces with toric divisors having
self-intersection $((n-1, -1, -1, - n, 0))$ with $n= 1, \ldots, 8,
12$.  There are 22 surfaces $S$ at $r = 3$, of which three are
non-toric.  The three non-toric $r = 3$ bases are those resulting from
blowing up the middle $-1$ curve in the chain of curves $-1, -1, -  n$
at a generic point, when $n= -1, -2, -3$.  (For $-4$ or below, this
construction gives intersecting curves $-2, -n$
with $n \leq -4$, which gives
rise to a $(4, 6)$ singularity at the intersection point.)
The non-toric example in Figure~\ref{f:blowups} is the example of this
type with $n= 1$.  Continuing further, at $r = 4$ there are a total of
49 distinct base topologies, of which 27 are toric, another 7 are
semi-toric but not toric, and another 15 are non-toric and cannot be
identified by toric or semi-toric analysis.  As $r$ increases, the
fraction of bases that are non-toric increases.

\begin{figure}
\centering
\includegraphics[height=9cm]{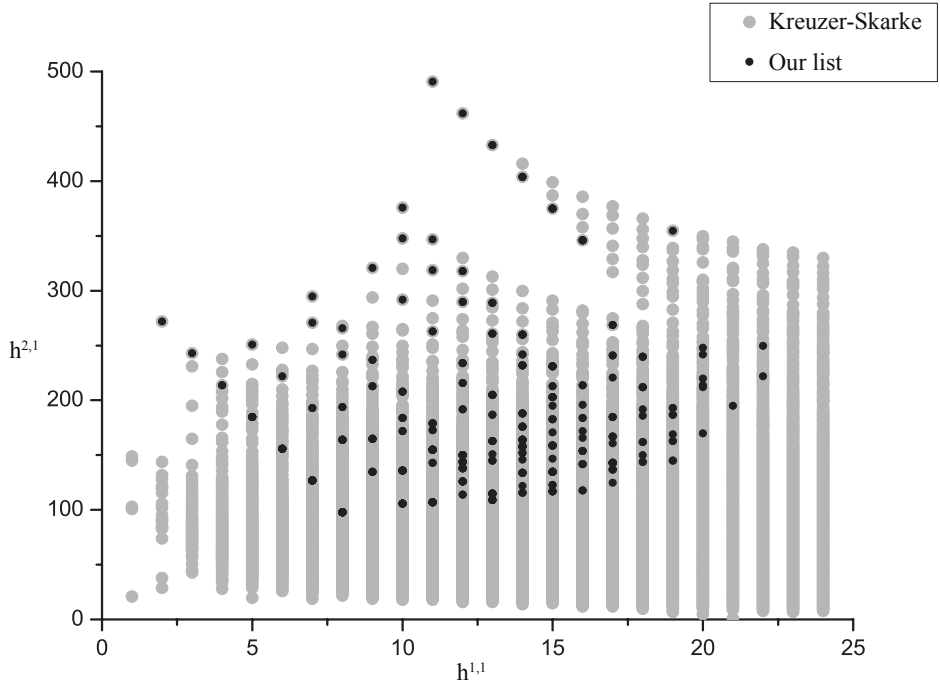}
\caption[x]{\footnotesize The Hodge numbers of  the generic EFS
(elliptically fibered with section) Calabi-Yau threefolds
over all smooth complex surface
  bases with rk(Pic$(S))<8$  are represented by
  black dots. The 6D  theories that result from compactification on
  these CY3s have $T<7$  tensor supermultiplets. The Hodge
  numbers in the Kreuzer-Skarke database are represented by gray
  dots. }\label{f:hodgeT7}
\end{figure}

We plot the Hodge numbers of the generic elliptically fibered
Calabi-Yau threefolds over the bases in this set in Figure
\ref{f:hodgeT7}. We also plot the Hodge numbers in  the Kreuzer-Skarke
database \cite{Kreuzer-Skarke}.  In principle, any elliptically
fibered Calabi-Yau threefold with $h^{1, 1} (X) < 9$ should be
supported over one of the bases in this class.  Note that many of the
Hodge number pairs in this range can be realized by tuning Weierstrass
models over the bases we have identified to realize larger gauge
groups associated with enhanced Kodaira singularities, along the lines
of the analysis in \cite{JohnsonTaylor}.  For example, the
Kreuzer-Skarke list contains a Calabi-Yau threefold with Hodge numbers
$(3, 231)$.  This Calabi-Yau can be realized by tuning an $SU(2)$
gauge group on the divisor class $H$ in $\P^2$.  Similarly, the
Calabi-Yau with Hodge numbers $(3, 195)$ can be realized by tuning an
$SU(2)$ on a quadratic curve in the divisor class $2H$ in $\P^2$,
etc. Note, however, that this analysis guarantees that the
Calabi-Yau's with certain large Hodge numbers such as $(1, 149), (2,
144), \ldots$ cannot be realized as elliptic fibrations.

The list also contains all the generalized del Pezzo surfaces listed
in \cite{Derenthal} from degree 6 to degree 3. The algorithm presented
in this paper is a potential tool to generate all the generalized del
Pezzo surfaces. The number of (-1)-curves on generalized del Pezzo
surfaces are listed in \cite{Derenthal}. These numbers exactly agree
with the numbers generated with our method in Proposition
\ref{p:gcurves}. For instance, consider $r=6$, if
Sing$(S)=\{(1,-1,-1,-1,0,0,0)\}$ or $\{0,1,-1,0,0,0,0\}$, then the
number of $(-1)$-curves is exactly 21, which can be found in Table 7
of \cite{Derenthal}. We can also run this algorithm for generalized
del Pezzo surfaces
up to $r=7, 8$,
restricting to bases without $-3$ curves or below. For $r = 7,$
when
the case where three curves all meet at a point is ignored, we get 46
different intersection structures, matching the results of
\cite{Derenthal}. When one allows three curves to meet 
at a point, a new base with 7 (-2)-curves that do not intersect each
other appears. This base is labelled by $7\mathbf{A}_1$ in the
literature, and it only occurs over a field of characteristic 2. From
our point of view, this base is forbidden by a geometric
obstruction that is not apparent from the combinatorics of the
effective cone.
The complete list of all 47 $r = 8$ bases including the $7 {\bf A}_1$
base is given in \cite{Urabe}.
Similarly, for $r = 8$, the algorithm when triple intersections (but
not special blow-ups) are allowed produces 76 bases, including the
74 identified in \cite{Derenthal} and two others that appear in
\cite{Urabe} that have geometric obstructions over $\C$.  The list in
\cite{Urabe} also contains one further base with $8 {\bf A}_1$, which
is not identified by our algorithm and also has a geometric
obstruction over $\C$.

When applied to bases with rk(Pic$(S))=9$, we might hope to use this
approach to reproduce the complete list of types of rational elliptic
surfaces with degenerate elliptic fibers as divisors, {\it i.e., Persson's list} \cite{Persson}\cite{Miranda}. 
This would provide a simple context in which to attempt to
systematically address the issues of multiply intersecting curves,
special blowups, and potentially infinite numbers of generators for
the effective cone.

\section{Classification of all bases for  elliptic
  Calabi-Yau threefolds with
  $h^{2,1}\geq 150$}
\label{sec:large}

Another set of bases that we can completely classify is the set of
bases that support elliptic Calabi-Yau threefolds $X$ with
$h^{2,1}(X)\geq 150$. This subset is relatively easy to study, because
it turns out that the difference between $|\mathrm{Neg}(S)|$ and
rk(Pic$(S))$ is small. Hence the dual cone problem is easier to solve,
though still computationally expensive. Also, the situations in which
three negative curves intersect each other or the dual of Neg$(S)$
contains a negative curve never happen, nor does an infinite number of
negative curves ever arise.  The only issue that could in principle
make our list incomplete is the presence of special blow-ups. Recall
that special blow-ups only happen when $S'$ contains a curve with
self-intersection $(-2)$ or lower, and the last entry in its vector
representation is (-2) or lower.  We begin by discussing the results
of our analysis, and then in subsection \ref{sec:special} we explain
why special blow-ups can essentially be ruled out for this class of
bases.  The analysis described in \S\ref{sec:special} also makes it
clear that in the range $h^{2, 1}\geq 150$ we do not expect higher
degree $-1$ curves, so that it should not be necessary to check the
dual cone.  We nonetheless explicitly checked the dual cone in the
cases where $h^{2, 1}\geq 200$ and confirmed that in this range the
effective cone is automatically produced correctly by $q$-operations
on low-degree curves, and does not miss
any higher degree $-1$ curves.  We did not explicitly check this in
the range $150 \leq h^{2, 1} < 200$, due to the computational time
  needed, but we expect from the arguments in \S\ref{sec:special} that
  the results in that range would be the same as those for $h^{2, 1}
  \geq 200$.

\subsection{Statistics of bases giving generic EFS CY3's
with large $h^{2,1}$}

In total our algorithm produces 6511 bases over which the generic
elliptic Calabi-Yau threefold has $h^{2, 1} (X) \geq 150$.  This set
of bases includes $\mathbb{P}^2$ and $\mathbb{F}_n$, with $n=0\sim 8$
or 12.  All the 3871 toric and  toric +
semi-toric bases identified in
\cite{Toric, semi-toric} giving threefolds with $h^{2,1}\geq 150$ can
be found in our list (including ones generated from blowing up
-9/-10/-11 curves at generic points). Hence the number of new bases
which is not in the list of \cite{semi-toric} is 2640.

\begin{figure}
\centering
\includegraphics[height=9cm]{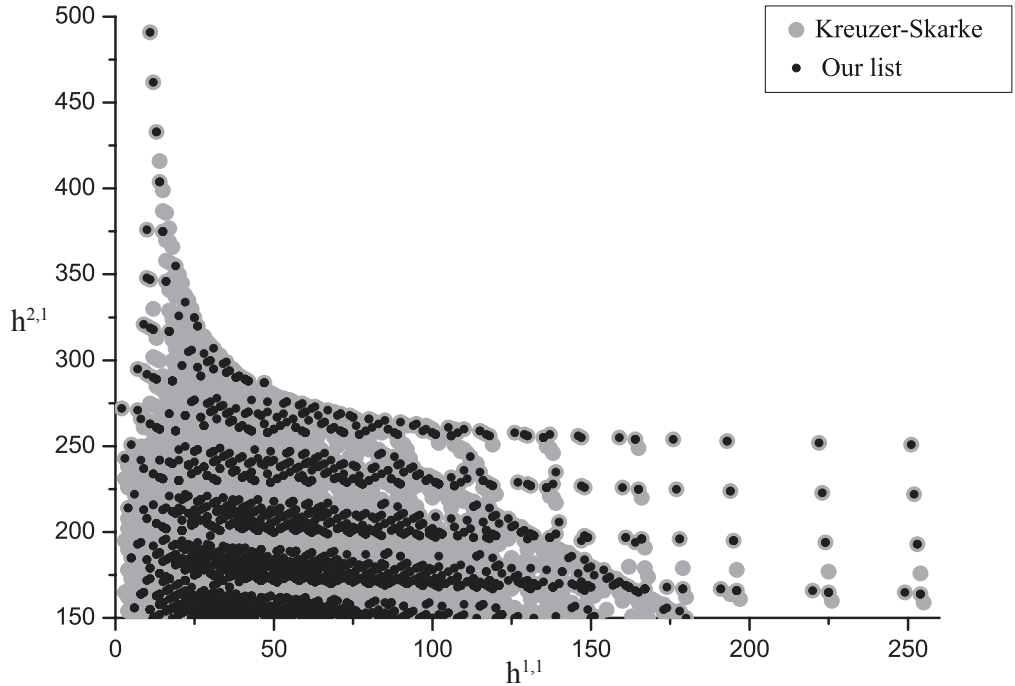}
\caption[x]{\footnotesize The Hodge numbers of all the EFS Calabi-Yau
  threefolds with $h^{2,1}\geq 150$ generated by generic elliptic
  fibrations over smooth surfaces are represented by black dots. The
  Hodge numbers in the Kreuzer-Skarke database are represented by gray
  dots.}\label{f:hodge150}
\end{figure}

We plot the Hodge numbers of the generic elliptic Calabi-Yau
threefolds over this set of bases in Figure \ref{f:hodge150}.  Generic
elliptic fibrations over the 6511 distinct bases give CY threefolds
with 1278 distinct Hodge number pairs.  For
very large $h^{2, 1}$, as discussed in \cite{JohnsonTaylor}, all known
Calabi-Yau threefolds can be realized as elliptic fibrations over
toric or semi-toric bases.  The first non-toric base arises from a
Calabi-Yau with Hodge numbers $(19, 355)$; these Hodge numbers can
also, however, be found from a Calabi-Yau that is elliptically fibered
over a toric base, where the toric base is a limit of the non-toric
base containing an extra $-2$ curve.  Similar situations arise at
other large values of $h^{2, 1}$.  The largest value of $h^{2, 1}$ at
which there is a non-toric base with no toric equivalent is at the
Hodge numbers $(29, 299)$.  In the following subsection we describe
this base, as well as other new non-toric bases that we have
identified with generic elliptic fibrations having Hodge numbers that
are not in the Kreuzer Skarke database.

\begin{figure}
\begin{center}
\includegraphics[width=10cm]{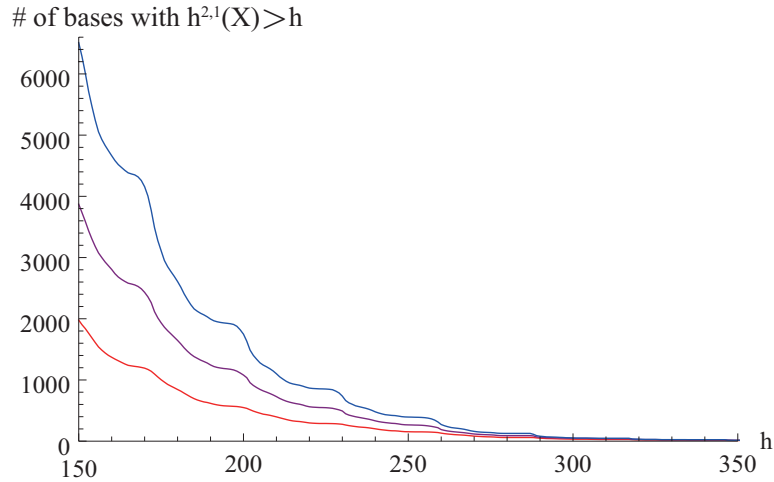}
\end{center}
\caption[x]{\footnotesize The number of toric (red), semi-toric
  (purple), and completely general including non-toric (blue) bases
  over which the generic elliptic Calabi-Yau threefold has $h^{2,
    1}(X)\geq h$ for a range of values of $h$.}
\label{f:large}
\end{figure}

A graph of the number of toric, semi-toric, and non-toric (meaning
neither toric nor semi-toric) bases with Hodge numbers exceeding
particular values of $h^{2, 1}$ is shown in Figure~\ref{f:large}.
It is notable that the
number of  non-toric constructions is not dramatically larger than the
number of toric bases anywhere in the range considered.  In principle,
one might have imagined that toric constructions would only represent
a very specialized class of Calabi-Yau threefolds.  At least for
elliptic CY threefolds with $h^{2, 1} \geq 150$, we see that this is not
the case.  In fact, toric geometry seems to be surprisingly effective
at constructing a representative sample of Calabi-Yau threefolds at
large Hodge numbers.

\subsection{New Calabi-Yau threefolds}

We have identified 15 bases that give rise to elliptic Calabi-Yau
threefolds with Hodge numbers that are not in the Kreuzer-Skarke
database. Their Hodge numbers are:
\be
\bsp
(h^{1,1},h^{2,1})\ =\ &(29, 299), (48, 270), (30, 270), (59, 269), (41, 269),(70, 268),\\
& (31, 241), (31, 241), (66, 240), (42, 240), (89, 239),(20, 214),\\
 & (84, 210), (149, 179), (104, 152)\,.
\end{split}
\ee
The mirrors of these Calabi-Yau threefolds also do not appear in the
Kreuzer-Skarke database, so this represents 30 new Calabi-Yau
threefolds.
The fact that only 15 of the 1278 Hodge number pairs produced by our
algorithm represent Calabi-Yau threefolds with Hodge numbers that are
not produced through toric constructions suggests that toric methods
are remarkably effective in producing a representative sample of
Calabi-Yau manifolds, at least for elliptic threefolds with large
Hodge numbers.
Note that while 8 semi-toric bases were identified in \cite{semi-toric}
that give rise to elliptic Calabi-Yau threefolds with Hodge numbers
not included in the Kreuzer-Skarke database, the largest Hodge number 
$h^{2, 1}$
for any of these was $h^{2, 1} = 31$.

The new Calabi-Yau threefold with the largest value of $h^{2, 1}$,
which has Hodge numbers $(29, 299)$, can be understood by explicitly
analyzing the base geometry.  This geometry is closely related to the
set of bases studied in \cite{JohnsonTaylor}.  If we begin with the
toric base having self-intersections
\begin{equation}
 ((-12, -1, -2, -2, -3, -1, -4, -1, -3, -1, 6, 0)) \,,
\end{equation}
which is associated with a Calabi-Yau having Hodge numbers $(28,
304)$, and then blow up a generic point on the $-4$ curve, we get a
non-toric base having no toric equivalent, over which the generic
Calabi-Yau elliptic fibration has Hodge numbers $(29, 299)$.  The
shift by $-5$ in $h^{2, 1}$ arises from a shift of $-29$ from the
blowup, and $+24$ from the difference in the contribution to $V$ from
the dimensions of the groups $SO(8), F_4$.
By explicitly analyzing possible sequences of blowups on a single fiber, it can
be shown fairly readily that this base is the first one encountered as
$h^{2, 1}$ decreases that is non-toric and is not equivalent to a
toric base.  Note that after the non-toric blowup  it is not possible to
blow up another point on the new $-5$ curve, since the non-Higgsable
cluster $-6$ cannot be adjacent to a $-3, -2, -2$ cluster
\cite{Classbasis}.
One can blow up the point between the $-1$ and $6$ curves, giving a
new base with generic elliptic Calabi-Yau having Hodge numbers $(31,
271)$.  This non-toric base is not equivalent to a toric base, but has
the same Hodge numbers as a  closely related toric base.  Blowing up a
generic point on the $(29, 299)$ base gives a base with Hodge numbers
$(30, 270)$, also on the list above.  And the base with Hodge numbers
$(48, 270)$ can be constructed by starting with a toric base having
self-intersections $((-12, -1, -2, -2, -3, -1, -5, -1, -3, -2, -1, -8,
-1, -2, -2, -2, 7))$ and blowing up a point on the middle $-2$ curve
in the last $A_3$ sequence.  The rest of the new bases and associated
Calabi-Yau threefolds can be understood in a similar fashion.

Note that all of these non-toric constructions are fairly close to
toric bases, consisting for the most part of long linear chains of
non-Higgsable clusters connected by $-1$ curves, with occasional
adornments that take the bases outside the toric class.   While we
expect that at smaller $h^{2, 1}$ the intersection structure of
non-toric bases becomes more complex, the general picture that we see
at large $h^{2, 1}$ matches well with the results of \cite{semi-toric},
and suggest that going beyond toric bases, and including branching and
loops in the intersection structure of the base does not dramatically
increase the complexity or number of possible base surfaces that
can support elliptic Calabi-Yau threefolds.

\subsection{Special blow-ups}
\label{sec:special}

We conclude this section with a brief description of why special
blow-ups do not occur in this set of bases associated with elliptic
Calabi-Yau's having large $h^{2, 1}$.  While we do not have a formal
mathematical proof that this cannot occur, we argue below that special
blowups cannot arise for any base giving a generic elliptic fibration
with $h^{2, 1} > 135$.  Related arguments explain why at large
$h^{2, 1}$ there are no issues with multiple intersecting negative
curves, or infinitely generated effective cones.

In the full set of data that we generated without including the
possibility of special blow-ups, it turns out that there is never a
negative curve with a (-2) in the vector representation; in fact there
is not even a single negative curve with degree 2 or higher in any of
the bases. The reason is   essentially that curves of high degree are
strongly constrained in the presence of non-Higgsable clusters.
For instance, when we consider the bases that
come from blowing up $\mathbb{F}_{12}$, the (-12)-curve
$(-5,6,-1,0,\dots)$ tightly restricts the possible form of negative
curves. To ensure that a negative curve intersects non-negatively with
the (-12)-curve, all the curves with degree 2-5 are excluded, and the
first  valid
example of higher degree is $(6,-5,0,\dots)$. In our data it turns out
that there is never a (-1)-curve in this form. 
This is due to the fact that
we only keep bases with large $h^{2,1}(X)$. 
The same features that obstruct the appearance of such a (-1)-curve
even more strongly obstruct the appearance of a (-2)-curve or below
that could arise from a special blowup..

To be more explicit, let us attempt to identify the base with maximal
$h^{2,1}(X)$ that arises as a blowup of $\F_{12}$, and which contains
the (-1)-curve $(6,-5,0,(-1)\times 12)$.  This is the first
$(-1)$-curve of higher degree that can appear on a blowup of
$\F_{12}$.  We claim that the base of interest is just
$\mathbb{F}_{12}$ blown up at 13 generic points, denoted by
Bl$_{13}\mathbb{F}_{12}$, with $h^{2,1}=491-13\times 29=114$.  This
base has no NHC aside from the original (-12) curve.  It is not
possible to keep the same degree 6 (-1) curve and increase $h^{2, 1}$
by choosing specific points to generate additional nonabelian groups,
since, for example, a (-3) curve that could support an $SU(3)$ factor
would need to be on a curve of the form {\it e.g.}  $(0, 0, 0, 1, -1,
-1, 0, \ldots, 0)$, which cannot happen as it would have negative
intersection with the degree 6 negative curve.  Further non-Abelian
non-Higgsable gauge groups can be added by further blowups, but this
procedure will always lower $h^{2,1}(X)$. One may wonder if there is a
base $S$ that contains a (-1)-curve like $C=(6,-5,0,(-1)\times
m,0,(-1)\times (12-m))$ with $h^{2,1}(X)$ greater than 114. This is
impossible, because one can permute the last (-1)-entry with this
0-entry in the middle. For other curves on $S$, this permutation is
always feasible except for the curve of the form
$(0,0,\dots,1,\dots,-1)$, where the 1-entry is going to be
permuted. However this curve is not compatible with $C$, because the
intersection number between them is negative. After the permutation,
the base becomes a blow up of Bl$_{13}\mathbb{F}_{12}$, hence its
$h^{2,1}(X)$ is smaller than 114.
This shows that a degree 6 curve of self-intersection -1 cannot arise
from blowups of $\F_{12}$ for any base associated with an elliptic
Calabi-Yau having $h^{2, 1} > 114$.

Now considering other type of curves with degree higher than 1 on
blowups of $\F_{12}$, the base with maximal $h^{2,1}(X)$ that contains
the curve is also the base with smallest Picard rank (or length of
vector representation). For example, the curve \be
D=(12,-10,0,(-2)\times 9,(-1)\times 6,-2) \ee is a curve that can be
generated by special blow-up. The maximal NHC that is compatible with
this $D$ are three (-3)-curves of the form
$(0,\dots,1,\dots,-1,-1,\dots)$. Hence the maximal
$h^{2,1}(X)=491-17\times 29+3\times 8=22$. In general the maximal
$h^{2,1}(X)$ of bases containing these types of curves which are generated by special
blow-up is even less than 114, hence in our regime they will never
appear.  This conclusion seems quite clear from the form of curves of
degree $6, 12, 18, \ldots$, though we have not attempted a systematic and
complete mathematical analysis of all possibilities.

Similarly for the bases that come from blowing up $\mathbb{F}_8$,
which contains a (-8)-curve $(-3,4,-1,0,\dots)$, the lowest degree
negative curve with degree greater than 1 is $(4,-3,0,(-1)\times
8)$. The maximal $h^{2,1}(X)=376-9\times 29=115$. We list these
numbers for each $\mathbb{F}_n$ in Table \ref{t:h21}; the largest
value in that table is $135$ for $\mathbb{F}_3$. The conclusion
is that  special blow-ups should not bother us in the regime $h^{2,1}(X)> 135$.
\begin{table}
\centering
\begin{tabular}{|c|c|c|c|c|c|c|c|c|}
\hline
$n$&0,1,2&3&4&5&6&7&8&12\\
\hline
max$(h^{2,1}(X))$&127&135&126&121&118&116&115&114\\
\hline
\end{tabular}
\caption[x]{\footnotesize The maximal $h^{2,1}(X)$ of bases that come from blowing up $\mathbb{F}_n$, and have a negative curve with degree higher than 1.}\label{t:h21}
\end{table}

The absence of negative curves of degree higher than one helps to
explain why there are no problems with configurations of multiple
mutually intersecting curves, or infinitely generated effective cones,
when $h^{2, 1}$ is reasonably large.  For example, for bases that
arise as blowups of $\F_{12}$, and on which the only $(-1)$ curves are
of degree 1, such curves always take the form $(1, -1, 0, \ldots, 0,
-1, 0, \ldots, 0)$ or $(0, \ldots, 0, 1, 0, \ldots, 0)$.  It is easy
to see that there cannot be three curves in these forms that all
intersect pairwise.  Furthermore, there are clearly a finite number of
negative curves on any such base.  While the constraints are weaker
for bases that arise as blowups of $\F_m$ for small $m$, the only
surfaces without curves of self-intersection -3 or below are the
generalized del Pezzo surfaces, and we know that the first appearance
of three mutual intersecting curves in that context occurs on $dP_6$,
associated with a Calabi-Yau having $h^{2, 1} = 98$.

Thus, it seems in principle that it should be possible to carry out
the systematic analysis of bases as far down as roughly $h^{2, 1} \sim
135$ before any of these issues will arise.  For bases without
higher-degree negative curves, the algorithm can also be made more
efficient since it is not necessary to check the dual cone condition.
The remaining issue which this does not address, however, is whether
$- K$ is in the effective cone for all such bases.

\section{Conclusions}
\label{sec:conclusions} 

In this paper we have developed a combinatorial approach to exploring
the set of smooth non-toric base surfaces that can support
elliptically fibered Calabi-Yau threefolds.  This approach has enabled
us to carry out a systematic enumeration of base surfaces with small
Picard number and base surfaces that support elliptic Calabi-Yau
manifolds $X$ with large Hodge number $h^{2, 1}(X) \geq 150$.  

There are a few issues that make it difficult to extend the
enumeration carried out here to the complete set of non-toric bases
associated with arbitrarily small values of $h^{2, 1} (X)$.
In particular, the combinatorial data of the cone of effective
divisors  in $\Z^{1, T}$
that we use here to characterize
surfaces does not capture all geometric information;  in some cases
there are  surfaces ``of the same type'' that share this combinatorial
description but have different more detailed structure.  Related to this,
in some cases there are ``special blow-ups'' that produce curves of
self-intersection -2 or below that are difficult to systematically
detect.  Neither of these issues arises in the specific classes of
surfaces that we have constructed here, but they would need to be
dealt with for a complete enumeration of bases, or  of elliptic Calabi-Yau
threefolds.  There is also the issue that the number of generators of
the effective cone can be infinite; this seems to only occur, however,
at limiting bases that cannot be blown up further to give further
valid bases for elliptic CY fibrations.

One of the main lessons of this work is that including completely
general non-toric bases does not seem to dramatically expand the range
of possible constructions for elliptic Calabi-Yau threefolds, at least
at large Hodge number.  This extends further the qualitative results
of \cite{semi-toric}, which showed that adding nontrivial branching and
loops to the intersection structure of negative self-intersection
curves in the base does not dramatically increase complexity.  Indeed,
particularly at large Hodge numbers it seems that toric  bases do a
remarkably good job of characterizing the set of possible  bases for
elliptic Calabi-Yau threefolds, at least at a qualitative level, and
that while adding bases with semi-toric or non-toric structure
increases the number of possibilities, it does not do so by many
orders of magnitude.  

The bases constructed here could be used in principle to construct all
possible elliptic Calabi-Yau threefolds with $h^{2, 1} \geq 150$, by
including all tunings of additional nonabelian gauge groups, matter
representations at codimension two, and abelian gauge groups, along
the lines of \cite{JohnsonTaylor}.  While this would be a
computationally intensive endeavor, and there are still some open
questions regarding possible matter representations and codimension
two singularities, much of the understanding is in place to make this
a tractable proposition.  Recent work from a number of different
directions \cite{WT-Hodge, Candelas-c-Skarke, JohnsonTaylor, Gray-hl}
suggests that the majority of known Calabi-Yau threefolds are
elliptic, particularly those with large Hodge numbers.
It is interesting to speculate that the set of elliptic CY threefolds
that could be constructed using the bases determined here could in
fact comprise all, or almost all, of the Calabi-Yau manifolds with
$h^{2, 1}\geq 150$, independent of the elliptically fibered condition.
Of course, more insight into the structure of general Calabi-Yau
threefolds would be necessary to verify any such claim.

Some of the issues encountered in this paper pose interesting
questions for the program of ``string universality'' in six
dimensions.  It is known that all quantum-consistent massless spectra for 10-dimensional
theories of gravity with minimal supersymmetry are realized in string
theory \cite{10D-universality}, and it has been conjectured that a
similar statement is true in six dimensions \cite{6D-universality}.
The close connection between the geometry of F-theory constructions
and the structure of the corresponding low-energy six-dimensional
supergravity theory was used in \cite{KMT, KMT-II} to relate
constraints between these two pictures.  In particular, the
intersection structure on the F-theory base maps to and constrains the
structure of the dyonic string lattice in the 6D theory
\cite{Seiberg-Taylor}.  From this point of view, the set of F-theory
vacua on Calabi-Yau manifolds with large $h^{2, 1}$, corresponding in
the 6D supergravity picture to theories with large numbers of neutral
scalar fields, may be a natural context in which to strengthen the
evidence for string universality in 6D.  For example, it would be
interesting to try to demonstrate that the only consistent low-energy
6D supergravity theories with one supersymmetry and more than 150
neutral scalar fields are those that arise from F-theory constructions
on the bases we have constructed here.  One interesting challenge for
extrapolating such a program to lower $h^{2, 1}$ is to find a low
energy understanding of the different types of constraints discussed
in \S\ref{sec:obstructions} associated with {\it e.g.}  Pappus's
theorem, or the configuration described in Table~\ref{t:4I3}, each of
which indicates that a certain structure of effective cone for the 6D
supersymmetric string charge lattice that may appear consistent,
cannot be realized in F-theory by  a sensible base geometry.  Such issues
may provide a useful insight into the precise correspondence
between UV consistency conditions on low-energy six-dimensional
supergravity theories and the geometry of F-theory constructions.

The datasets described in Sections \ref{sec:small} and \ref{sec:large}
are available online at \cite{online-data}.
\vspace*{0.1in}

\section*{Acknowledgements}
We would like to thank Lara Anderson, James
Gray, Sam Johnson, Gabriella Martini, and David Morrison for helpful
discussions.  This research was supported by the DOE under contract 
\#DE-SC00012567.

\address{Center for Theoretical Physics,\\Department of Physics\\Massachusetts Institute of Technology\\77 Massachusetts Avenue\\Cambridge, MA 02139, USA}

\email{{\tt wati} {\rm at} {\tt mit.edu}}
\email{{\tt wangyn} {\rm at} {\tt mit.edu}}

\end{document}